\documentclass[usenatbib,usegraphicx]{mn2e}
\usepackage{amssymb}
\newcommand{\halfspace}{\hspace{1pt}}

\newcommand\diff{{\rm d}}
\newcommand\dnu{{\rm d}\nu}

\newcommand{\kms}{\mathop{\rm km \ s^{-1}\,}\nolimits}
\newcommand{\Lya}{Ly$\alpha$}

\newcommand\HI{{\hbox{H\halfspace$\rm \scriptstyle I$}}}

\newcommand\HeII{{\hbox{He\halfspace$\rm \scriptstyle II$}}}
\newcommand\HeIII{{\hbox{He\halfspace$\rm \scriptstyle III$}}}

\newcommand\lsim{~\lower.5ex\hbox{$\buildrel < \over \sim$}~}
\newcommand\gsim{~\lower.5ex\hbox{$\buildrel > \over \sim$}~}

\title[The impact of helium reionization on the structure of the intergalactic medium]{The impact of helium reionization on the structure of the intergalactic medium}
       
\author[Avery Meiksin, Eric R. Tittley]{
        Avery Meiksin$^{1}$\thanks{E-mail:\ A.Meiksin@ed.ac.uk (AM)},
        Eric R. Tittley$^{1}$\\
        $^{1}$SUPA\thanks{Scottish Universities Physics Alliance},
	Institute for Astronomy, University of Edinburgh,
        Blackford Hill, Edinburgh\ EH9\ 3HJ, UK}

\shortcites{}

\begin{document}

\date{Accepted . Received ; in original form }
\pagerange{\pageref{firstpage}--\pageref{lastpage}} \pubyear{2011}
\maketitle
\label{firstpage}

\begin{abstract}

  We examine the impact of helium reionization on the structure of the
  intergalactic medium (IGM). We model the reionization using a
  radiative transfer (RT) code coupled to the combined gravity
  hydrodynamics code {\texttt{Enzo}}. Neutral hydrogen and helium are
  initially ionized by a starburst spectrum, which is allowed to
  gradually evolve into a power law spectrum over the redshift
  interval $3.0<z<4.0$. The temperature-density relation of the gas is
  found to fan out and flatten following \HeII\ reionization, with an
  inversion for high overdensities of $\rho/\langle\rho\rangle>5$.
  Peculiar velocities of up to $10\kms$ are induced by the increased
  pressure, with the gas density field distorted over large coherent
  regions by 10--20 per cent., and the dark matter by levels of 1 per
  cent. The photoionization-induced flows may thus distort the matter
  power spectrum at comoving wavenumbers $k>0.5\,h\,{\rm Mpc}^{-1}$ by
  a few per cent. by $z=2$.

  Absorption spectra for \HI\ and \HeII\ are drawn from the
  simulations, and absorption lines are fit to the spectra. The
  increased temperature following \HeII\ reionization reduces the line
  centre optical depths, resulting in an enhancement in the fraction
  of very low optical depth pixels and an overall distortion in the
  pixel flux distribution compared with \HeII\ reionization in the
  optically thin limit. A median Doppler parameter of $35\kms$ is
  obtained for the \HI\ absorption systems at $z=3$. Dividing into
  subsamples optically thick and optically thin at line centre reveals
  that the optically thick systems undergo only mild evolution while
  the optically thin systems evolve rapidly following \HeII\
  reionization. A comparison between \HeII\ and \HI\ absorption
  features shows a broad distribution in the \HeII\ and \HI\ column
  density ratio, peaking near the measured value and only slightly
  narrower than measured. The distribution of the ratio of \HeII\ and
  \HI\ Doppler parameters peaks midway between the thermally broadened
  and velocity broadened limits. A comparison with approximate
  simulation methods based on either a pseudo-hydrodynamical scheme or
  a non-RT hydrodynamical simulation with boosted \HeII\ heating rate
  shows moderately good agreement in the absorption line properties,
  but not to the precision to which they may be measured, and not over
  the full redshift range for which the high redshift \Lya\ forest is
  observed.

\end{abstract}


\section{Introduction}
\label{sec:Intro}

Cosmological simulations of the intergalactic medium (IGM) in the
context of cold dark matter (CDM) theories for structure formation
have proven extremely successful in reproducing many of the observed
properties of the IGM as measured through the \Lya\ forest. The mean
\HI\ transmission through the IGM is recovered for an \HI\ ionizing
background matching that of the measured Quasi-Stellar Object (QSO)
luminosity function for $z<3$, although an enhanced radiation field,
likely due to galaxies, is required at higher redshifts. Once
normalised to the mean transmission, the \HI\ column density
distribution is broadly recovered as well for a $\Lambda$CDM model, as
is the measured power spectrum of the \HI\ flux. See
\citet{2009RvMP...81.1405M} for a review.

Despite these successes, some of the statistics still show lack of
agreement with measured values. Prominent among these is the line
widths of the \HI\ absorption features, which tend to be broader than
the predictions \citep{2000ApJ...534...57B, 2000MNRAS.315..600T}. A
detailed comparison with measured spectra reveals that the
disagreement is among the optically thin \Lya\ systems, for which an
additional contribution of $10-15\kms$, added in quadrature to the
Doppler parameter, is required, while the widths of the optically
thick systems well match the measured values \citep{MBM01}. One
possible explanation is recent heat input at $z\lsim4$ into the
IGM. The additional required broadening corresponds to a temperature
increase of $17\times10^3$~K. The optically thin systems are
associated with moderate density or even underdense structures for
which the time to reach thermal equilibrium exceeds a Hubble time
\citep{Meiksin94, MR94}. The optically thin systems may thus retain a
memory of recent heat input at $z\lsim4$.

The most likely source of heating is \HeII\ reionization. The
abundance of QSOs suggests they would have ionized \HeII\ at a
redshift $3<z_{\HeII}<4.5$, although possibly as early as
$z_{\HeII}\simeq5$ for sufficiently hard spectra
\citep{Meiksin05}. Support for recent \HeII\ reionization has been
based on measurements of the \HeII\ \Lya\ optical depth of the
IGM. The measurements span the redshift range $2.2<z<3.8$ based on
{\it Hubble Space Telescope} ({\it HST}) observations of Q0302$-$003
($z_{\rm em}\simeq3.3$) \citep{1994Natur.370...35J,
  2000ApJ...534...69H}, {\it Hopkins Ultraviolet Telescope} and {\it
  Far Ultraviolet Spectroscopic Explorer} ({\it FUSE}) observations of
HS~1700$+$6416 ($z_{\rm em}\simeq2.7$) \citep{1996Natur.380...47D,
  2006A&A...455...91F}, {\it HST} and {\it FUSE} observations of
HE~2347$-$4342 ($z_{\rm em}\simeq2.9$) \citep{1997A&A...327..890R,
  2001Sci...293.1112K, 2002ApJ...564..542S, 2010arXiv1008.2957S}, and
{\it HST} observations of PKS~1935$-$692 ($z_{\rm em}\simeq3.2$)
\citep{1999AJ....117...56A}, SDSS J2346$-$0016 ($z_{\rm em}\simeq3.5$)
\citep{2004AJ....127..656Z,2008ApJ...686..195Z}, Q1157$+$3143 ($z_{\rm
  em}\simeq3.0$) \citep{Reimers05} and SDSS J1711$+$6052 ($z_{\rm
  em}\simeq3.8$) \citep{2008ApJ...686..195Z}. The \HeII\ \Lya\ optical
depths rise rapidly from $\tau_{\rm HeII}\simeq1.00\pm0.07$ averaged
over the redshift interval $z=2.2-2.6$ to $\tau_{\rm HeII}\simeq4.9$
at $z\simeq3.3$ \citep{2011ApJ...726..111S}, as if entering into the
epoch of \HeII\ reionization.

The origin of the UV background, however, is still not well
established. The high \HeII\ \Lya\ optical depths compared with the
corresponding values found for \HI\ generally requires that the UV
metagalactic background be soft between the \HeII\ and \HI\
photoelectric edges, with a ratio of \HI\ to \HeII\ photoionization
rates of $\Psi > 200$ at $z\approx3$ \citep{MM94}, corresponding to a
source spectral index for $f_\nu\sim \nu^{-\alpha_S}$ of
$\alpha_S>1.8$ after allowing for the filtering of the radiation
through the IGM \citep{MM94, HM96, BHVC06}. This is consistent with
the inferred spectral index of bright high redshift ($0.3<z<2.3$) QSOs
after correcting for IGM absorption \citep{2002ApJ...565..773T}, but
it is inconsistent with the much harder spectra found for dimmer QSOs
nearby ($z<0.67$) \citep{2004ApJ...615..135S}, for which minimal IGM
corrections are required. Possibly ionizing radiation from galaxies
boosts the \HI\ photoionization rate, however the contribution from
QSOs alone approximately matches the required rate to recover the
measured \HI\ \Lya\ optical depth \citep{Meiksin05,
  2005MNRAS.357.1178B}, leaving little room for additional sources.

The optical depth measurements are moreover found to fluctuate over a
wide range at a given redshift. This is most dramatically illustrated
by measurements of the \HeII\ \Lya\ forest compared with the
corresponding measurements of \HI. A wide spread is found for the
column density ratio $\eta=N_{\rm HeII}/N_{\rm HI}$, varying at least
over the range $4<\eta<600$ at $2.3<z<2.8$ \citep{Zheng04, Reimers05,
  2006A&A...455...91F, 2010arXiv1008.2957S}. The higher values
correspond to \HI\ to \HeII\ photoionization rate ratios of
$\Psi>1000$. Such fluctuations are expected during the epoch of \HeII\
reionization, so that the large \HeII\ \Lya\ optical depth values may
result from a mean over regions in which most of the helium is still
in the form of \HeII.

The interpretation is not unique, however. Due to the discreteness of
the sources of ionizing radiation and attenuation by the IGM, local
fluctuations are expected in the UV metagalactic background
\citep{1992MNRAS.258...36Z, 1998AJ....115.2206F, MW03}. The
distribution of fluctuations in $\eta$ measured in HE~2347$-$4342 are
well matched by a UV background dominated by QSOs with a soft spectral
index consistent with the data of \citet{2002ApJ...565..773T}
\citep{2004ApJ...600..570S, 1998AJ....115.2206F, BHVC06}, as are the
$\eta$ fluctuations measured in HS~1700$+$6416
\citep{2009RvMP...81.1405M}. Possibly a harder spectrum could be
accommodated in the presence of additional \HI\ ionizing sources other
than QSOs, but, again, there is not much room to add more sources
without over-ionizing the hydrogen. The full range of fluctuations in
the measured values of $\tau_{\rm HeII}$ in several QSO lines of sight
suggests the epoch of \HeII\ reionization may have completed as
recently as $z\simeq2.9$ \citep{2010ApJ...714..355F,
  2010arXiv1008.2957S}. By contrast, recent attempts to quantify the
thermal evolution of the IGM from the \HI\ \Lya\ forest measured in
QSO spectra indicate a rise in the temperature of gas at the mean
density for $z\simeq4$ \citep{2011MNRAS.410.1096B}, suggesting \HeII\
reionization was becoming widespread at these times. If both
interpretations are correct, then \HeII\ reionization was a drawn out
process extending over the redshift range $2.9\lsim z \lsim4.5$.

The purpose of this paper is to investigate the dynamical impact of
helium reionization on the structure of the intergalactic medium and
its observational signatures. While early simulations of the IGM
included helium reionization, they were done in the optically thin
limit. Cosmological simulations solving the radiative transfer (RT)
equation during helium reionization were restricted to the
gravitational component, which was used to scale the baryonic
component properties \citep{2007MNRAS.380.1369T,
  2009ApJ...694..842M}. In this paper, we implement radiation
hydrodynamics using the algorithm of \citet{BMW04} coupled to the
gravity-hydrodynamics code {\texttt{Enzo}}\footnote{Available from
  http://lca.ucsd.edu.} (specifically v.1.0.1). The hydrodynamical
response of the gas to the boost in heating when radiative transfer is
included could result in more rapid photo-evaporation of the gas in
small haloes, allowing reionization to occur more rapidly, as well as
in observational signatures on the \Lya\ forest such as increased line
broadening due to outflows from the haloes and a decrease in the ratio
of gas density to dark matter density. An increase in the gas
temperature will also alter the ionization fractions of hydrogen and
helium at a given gas density, producing modifications to the column
densities of the absorption features, the distribution function of
pixel fluxes, and the flux power spectrum, all basic statistics used
to quantify the \Lya\ forest and the predictions of cosmological
models for its structure.

All results are for a flat $\Lambda$CDM universe with the cosmological
parameters $\Omega_m=0.24$, $\Omega_bh^2=0.022$ and
$h=H_0/100~\kms=0.73$, representing the total mass density, baryon
density and Hubble constant, respectively. The power spectrum has
spectral index $n=0.95$, and is normalized to $\sigma_{8h^{-1}}=0.74$.

This paper is organized as follows. Estimates for the expected IGM
temperature boost following \HeII\ reionization are discussed in the
next section. The simulations are described in Sec.~3 and the results
presented in Sec.~4. The \HI\ and \HeII\ spectral signatures of \HeII\
reionization are presented in Sec.~5. A comparison with approximate
simulation methods is provided in Sec.~6. The principal conclusions
are summarised in the final section.

\section{Photoionization heating}
\label{sec:photoheating}

A radiation field with its specific energy density locally
approximated near the \HeII\ photoelectric edge by
$u_\nu=u_L(\nu/\nu_L)^{-\alpha}$ will photoionize the \HeII\ at the rate
per \HeII\ ion
\begin{equation}
\Gamma_{\HeII}=\int\,\dnu\,\frac{cu_\nu}{h\nu}\sigma_\nu
\simeq h^{-1}c\sigma_0\frac{u_L}{3+\alpha}
\label{eq:GiHeII}
\end{equation}
where $\nu_L$ is the frequency of the \HeII\ Lyman edge and the
photoionization cross-section is approximated as
$\sigma\simeq\sigma_0(\nu/\nu_L)^{-3}$.
The corresponding heating rate per \HeII\ ion is
\begin{eqnarray}
G_{\HeII}&=&\int\,\dnu\,\frac{cu_\nu}{h\nu}(h\nu-h\nu_L)\sigma_\nu\nonumber\\
&\simeq&c\sigma_0\nu_L\frac{u_L}{(2+\alpha)(3+\alpha)}.
\label{eq:GHeII}
\end{eqnarray}
The heating rate per ionization, normalized by the ionization
potential, is then
$\epsilon_{\HeII}=G_{\HeII}/(h\nu_L\Gamma_{\HeII})=1/(2+\alpha)$. The
energy injected into the gas is thus sensitive to the shape of the
local ionizing spectrum. A hardened radiation field with $\alpha<0$
can result in a large amount of energy deposited per ionization.

The spectrum is expected to harden within an ionization front because
hard photons are less likely to be absorbed than soft photons above
the photoeletric threshold for a given optical depth at the threshold
energy. In general, for a column density $N$ the optical depth above
the photoelectric threshold of hydrogen or singly ionized helium is
$\tau_\nu\simeq\tau_L(\nu/\nu_L)^{-3}$ where $\tau_L=\sigma_0N$. The
photoionization rate per atom/ion may be expressed in terms of the
incident radiation density from the source
$u^S_\nu=u^S_L(\nu/\nu_L)^{-\alpha_S}$ as
\begin{eqnarray}
\Gamma &=&
\int_{\nu_L}^\infty\,\dnu\,\exp\left[-\tau_L\left(\frac{\nu}{\nu_L}\right)^{-3}\right]\frac{cu^S_L\sigma_0}{h\nu}\left(\frac{\nu}{\nu_L}\right)^{-(3+\alpha_S)}\nonumber\\
&=&\frac{1}{3}\frac{cu^S_L}{hN}\tau_L^{-\frac{1}{3}\alpha_S}\gamma(1+\frac{1}{3}\alpha_S,\tau_L)
\label{eq:Gion}
\end{eqnarray}
where $\gamma(a,x)=\int_0^x\,dt\,e^{-t}t^{a-1}$ is the incomplete
gamma function. Similarly, the heating rate per atom/ion is
\begin{eqnarray}
G &=&
\int_{\nu_L}^\infty\,\dnu\,\exp\left[-\tau_L\left(\frac{\nu}{\nu_L}\right)^{-3}\right]\frac{cu^S_L\sigma_0}{h\nu}\left(\frac{\nu}{\nu_L}\right)^{-(3+\alpha_S)}\nonumber\\
&&\times(h\nu-h\nu_L)\nonumber\\
&=&\frac{1}{3}\frac{cu^S_L\nu_L}{N}\tau_L^{-\frac{1}{3}\alpha_S}\nonumber\\
&&\times\left[\tau_L^{1/3}\gamma(1+\frac{\alpha_S-1}{3},\tau_L)-\gamma(1+\frac{1}{3}\alpha_S,\tau_L)\right].
\label{eq:GHeat}
\end{eqnarray}

\begin{figure}
\scalebox{0.45}{\includegraphics{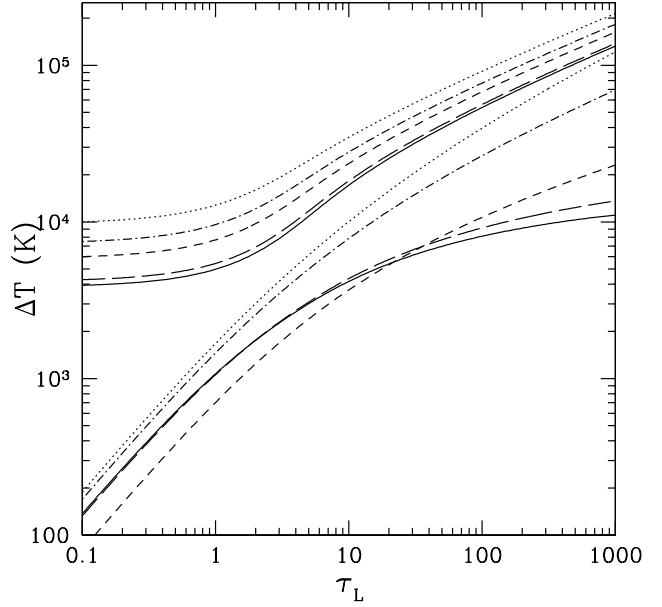}}
\caption{Temperature increment for ionizing \HeII\ to \HeIII\ as a
  function of the optical depth $\tau_L$ at the \HeII\ Lyman
  edge. Shown for a source spectrum $f_\nu\propto\nu^{-\alpha_S}$ with
  spectral index $\alpha_S=1.8$ (solid line), 1.5 (long-dashed line),
  0.5 (short-dashed line), 0 (dot-dashed line) and -0.5 (dotted
  line). The upper set of curves corresponds to reionization going to
  completion at a fixed optical depth, as would occur for radiation
  filtered through a foreground absorption system in ionization
  equilibrium. The temperature boost increases for harder source
  spectra. For large $\tau_L$, the temperature boost increases like
  $\tau_L^{1/3}$. The lower set of curves corresponds to the
  temperature boost for a time varying optical depth as an ionization
  front sweeps through a slab of material with an initial optical
  depth $\tau_L(0)$ given by the value $\tau_L$ indicated. Very high
  temperatures are reached for hard spectra.
}
\label{fig:temperature}
\end{figure}

The resulting heating rate per ionization of \HeII\ will produce a
temperature increment
\begin{equation}
\Delta T = \frac{2}{3}\frac{h\nu_L}{k_{\rm
    B}}\frac{\epsilon_{\HeII}}{2\frac{n_{\rm H}}{n_{\rm He}}+3}-\frac{T_i}{2\frac{n_{\rm H}}{n_{\rm He}}+3},
\label{eq:DeltaT}
\end{equation}
where $T_i$ is the initial gas temperature and
$\epsilon_{\HeII}=G/(h\nu_L\Gamma)$ for \HeII. The increment
corresponds to the complete ionization of \HeII\ to \HeIII\ at fixed
$\tau_L$, as would occur in the presence of an intervening Lyman limit
system, already in ionization equilibrium, lying between the source
and the region being photoionized if the Lyman limit system dominated
the optical depth at the Lyman edge. The radiation field reaching the
region being photoionized would be hardened by the absorption within
the intervening system. The temperature increment is shown as a
function of the optical depth $\tau_L$ at the \HeII\ Lyman edge for a
range of source spectral indices in Fig.~\ref{fig:temperature} (upper
set of curves). A helium mass fraction of $Y=0.248$ was adopted
\citep{2007ARNPS..57..463S}. A system with overdensity
$\rho/\langle\rho\rangle$ and comoving thickness $L$ would have an
optical depth at the \HeII\ Lyman edge of
\begin{equation}
  \tau_{L, {\rm
      HeII}}\simeq115\frac{\rho/\langle\rho\rangle}{100}\frac{(1+z)^2}{16}
  \frac{L}{1~{\rm Mpc}}.
\label{eq:tauLHeII}
\end{equation}
The radiation hardening through overdense systems with
$\rho/\langle\rho\rangle\sim200$ and comoving sizes of $0.05-0.25$~kpc
would produce temperature boosts in the ``shadows'' they cast in less
dense, more quickly photoionized structures, of up to $\Delta
T\simeq40000$~K, and even reaching $10^5$~K in the shadows of
intervening systems with $\tau_{L, {\rm HeII}}>100$. Radiative
recombination and excitation atomic line cooling will subsequently
lower the temperature of the gas once the ionization front has passed,
although the equilibrium temperature may never be achieved in rarefied
underdense regions because of the long required time scales
\citep{Meiksin94, MR94}.

Alternatively, the heating produced by the passage of an ionization
front through a slab of gas in which the optical depth decreases as
the ionization front sweeps through may be approximated by equating
the incident rate of ionizing photons with the rate at which the gas
in the slab is photoionized. For an initial optical depth $\tau_L(0)$
at the Lyman edge, the optical depth at the edge will evolve according
to
\begin{equation}
  \tau_L(t) = \tau_L(0)\left[1-k\frac{cu^S_L}{h}\frac{\sigma_0}{\tau_L(0)}t\right],
\label{eq:tauLt}
\end{equation}
where $k=1/\alpha_S$ for $\alpha_S>0$, $k=\log(\nu_{\rm max}/\nu_L)$
for $\alpha_S=0$ and $k=[(\nu_{\rm
  max}/\nu_L)^{-\alpha_S}-1]/(-\alpha_S)$ for $\alpha_S<0$, where
$\nu_{\rm max}$ is the maximum frequency for which photoionizing
photons are produced. Using Eqs.~(\ref{eq:Gion}) and (\ref{eq:GHeat})
for the photoionization and heating rates, the number density $n$ of
the species undergoing ionization and the thermal energy input per
unit volume $u$, as a function of $\tau_L$, are governed by
\begin{equation}
\frac{\diff\log
  n}{\diff\tau_L}=\frac{1}{3k\tau_L^{1+\alpha_S/3}}\gamma(1+\frac{\alpha_S}{3},\tau_L),
\label{eq:dlnndtauL}
\end{equation}
and
\begin{equation}
\frac{\diff u}{\diff\tau_L}=-\frac{1}{3}\frac{nh\nu_L}{k\tau_L^{1+\alpha_S/3}}\left[\tau_L^{1/3}\gamma(1+\frac{\alpha_S-1}{3},\tau_L)-\gamma(1+\frac{\alpha_S}{3},\tau_L)\right].
\label{eq:dudtauL}
\end{equation}
The resulting post-ionization temperature is shown in
Fig.~\ref{fig:temperature} (lower set of curves). Temperature boosts
of $\sim1000-10000$~K are typical, although hard spectra sources
($\alpha_S<0$) can produce substantially higher
temperatures. (Somewhat arbitrarily, $\nu_{\rm max}/\nu_L=4$ was
assumed for $\alpha_S\le0$.) Again, radiative recombination and
excitation line cooling will eventually decrease the gas temperature.

Temperature boosts following \HeII\ reionization typically ranging
between $\Delta T\simeq5000-20000$~K at $z=3$, but reaching as high as
$\Delta T\simeq30000-40000$~K directly after reionization, were found
in the reionization simulations of \citet{2007MNRAS.380.1369T} for a
source with spectral index $\alpha_S=0.5$, turning on after a
starburst-like spectrum photoionizes the hydrogen and neutral helium
to singly-ionized helium. For a pure power-law source spectrum
($\alpha_S=0.5$) and for a mini-quasar spectrum, lower overall
temperatures of $5000-20000$~K were obtained at $3<z<4$. Higher
temperatures were generally found in the denser
structures. Temperature boosts of $\Delta T\simeq10000-20000$~K were
found by \citet{2009ApJ...694..842M} for a QSO spectral distribution
with a mean spectral index of $\alpha_S=1.2$, and boosts of up to
$\Delta T\simeq30000$~K were found for harder source spectra with a
mean spectral index $\alpha_S=0.6$.

In addition to cooling by atomic processes, the gas will undergo
adiabatic cooling losses as it escapes from gravitational potential
wells that were able to bind the gas when cooler. The expansion
velocities will also broaden the absorption features
\citep{Meiksin94}. Previous \HeII\ reionization simulations, which
used the dark matter distribution to model the baryon density, were
not able to take these effects into account. In this paper, we take
the hydrodynamical response of the gas into account by coupling our
radiative transfer scheme to a gravity-hydrodynamics code.

\section{Simulations}

We have modified the combined hydrodynamics $N$-body gravity code
{\texttt{Enzo}} by replacing its ionization and atomic cooling modules
by ones employing the methods used by \citet{2007MNRAS.380.1369T}. In
summary, a probabilistic radiative transfer method is used to compute
the reionization of both hydrogen and helium. Atomic cooling from the
radiative recombination of the hydrogen and the two helium states is
included, along with collisional excitation of neutral hydrogen and
Compton cooling. Cooling due to the collisional ionization of hydrogen
and the collisional excitation and ionization of helium are negligible
for the temperatures encountered in the simulations.

For the hydrodynamical processes at the resolution of cosmological IGM
simulations, the dynamical time scales are much longer than the
radiative time scales in the vicinity of the ionization front. It would
be prohibitively expensive computationally and unnecessary to force
the hydrodynamics to evolve at the same time scale as the radiative
transfer. Hence, in our implementation the radiative transfer
time scale is allowed to drop below that of the hydrodynamics. The
energy equation is solved through operator splitting by computing the
ionization state of the gas and the amount of photoionization heating
and atomic cooling between the hydrodynamical time steps. During the
ionization timesteps, all parameters of the gas are taken to be those
at the end of the immediately preceding hydrodynamical timestep with
the exception of the fluid density and temperature. The density is
varied linearly from the previous hydrodynamical timestep to that of
the current ionization timestep while the radiative transfer processes
along with the gas temperature are evolved on their own, shorter,
adaptive time scale.

The simulations were carried out in a box $25h^{-1}\,{\rm Mpc}$
(comoving) on a side. This corresponds to the order of the expected
size of a \HeII\ ionization front at the helium reionization epoch,
when the characteristic diameter of the \HeIII\ zone produced by an
individual QSO is comparable to the mean distance between QSOs, which
is typically $\sim50$~Mpc (comoving) at $z\simeq3$
\citep{2003A&A...408..499W, Meiksin05}. Then {\texttt{Enzo}}
simulations were performed with $512^3$ gravitating particles and
$256^3$ grid zones for the fluid component, corresponding to a
resolution of 98~kpc (comoving), sufficient to resolve the Jeans
length.

The incident spectrum corresponds to a hybrid starburst--QSO model as
in \citet{2007MNRAS.380.1369T}:\ an initial starburst spectrum is
turned on at $z=8$ and transformed into a power-law spectrum
$f_\nu\propto\nu^{-0.5}$, turning on the power-law spectrum initially
at $z=4$ and ramping it up to its full value at $z=3$, while the
starburst spectrum is ramped down during the same interval.

\section{Results}
\subsection{Physical impact}

\begin{figure}
\scalebox{0.7}{\includegraphics{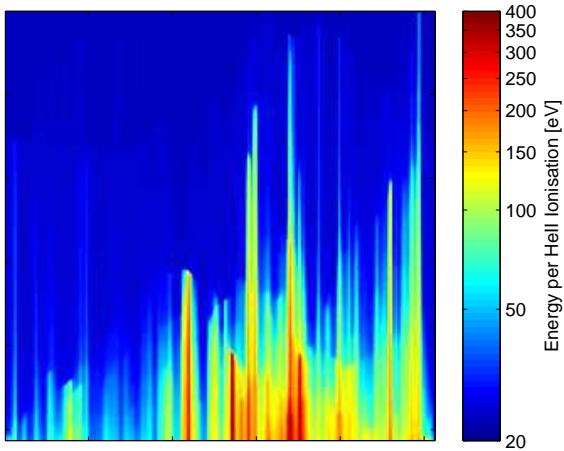}}
\caption{The heating rate per \HeII\ ionization at $z=3.3$ for a
  \HeII\ ionization front sweeping across the box from top to bottom.
}
\label{fig:epsHeII}
\end{figure}

The intensity of the starburst spectrum was adjusted to ensure the
hydrogen ionization front swept across the simulation volume by
$z\simeq6$, although some shadowed regions persist along the edge of
the box until $z\simeq5.5$. The intensity of the power-law spectrum
was adjusted so that the \HeII-ionization front swept across the box
by $z\simeq3.2$, again with some shadowing persisting to
$z\lsim3$. This corresponds to a hydrogen ionization rate from the QSO
at full intensity half that of the starburst at its full intensity.

The heating rate per ionization of \HeII\ as the \HeII\ ionization
front passes through the box at $z=3.3$ is shown in
Fig.~\ref{fig:epsHeII}. The filtering effect of the IGM in hardening
the ionizing radiation field is clearly discernible in the pre-ionized
gas ahead of the ionization front. The heating rate per ionization is
substantially enhanced, reaching values of up to $100-400$~eV. By
comparison, once the front has passed, the energy per \HeII\
ionization is $\sim22$~eV.

\begin{figure*}
\scalebox{0.9}{\includegraphics{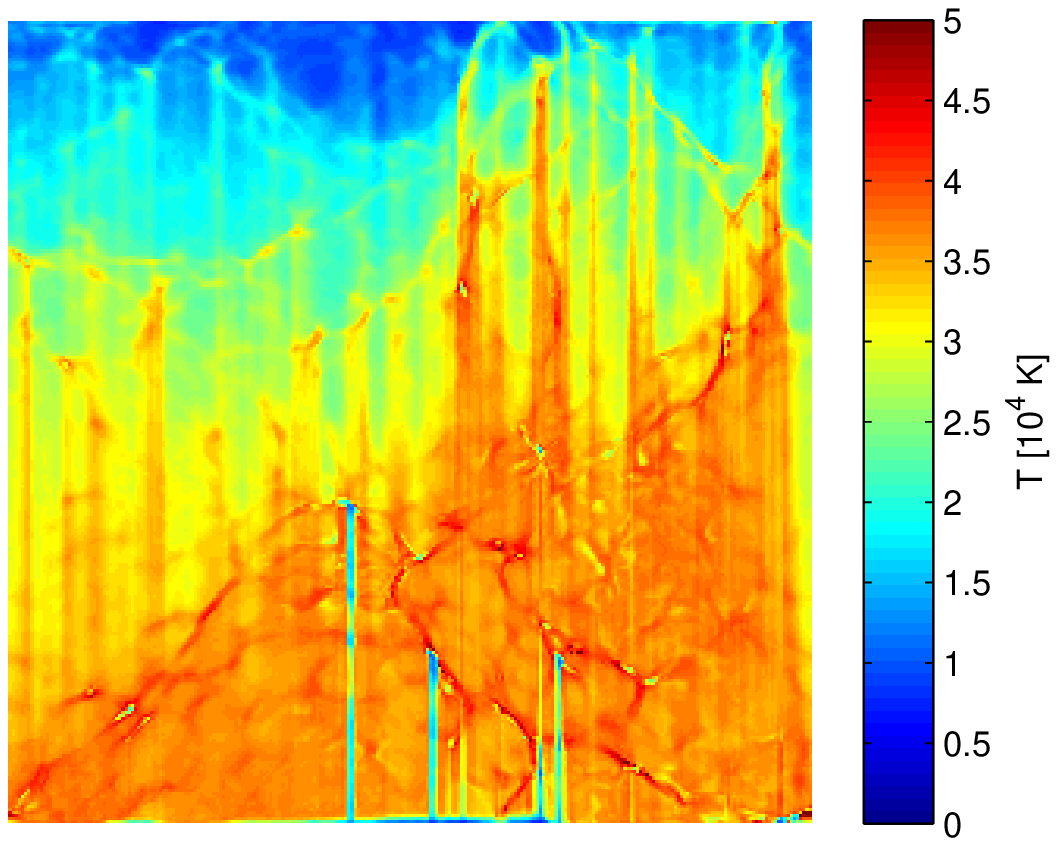}\includegraphics{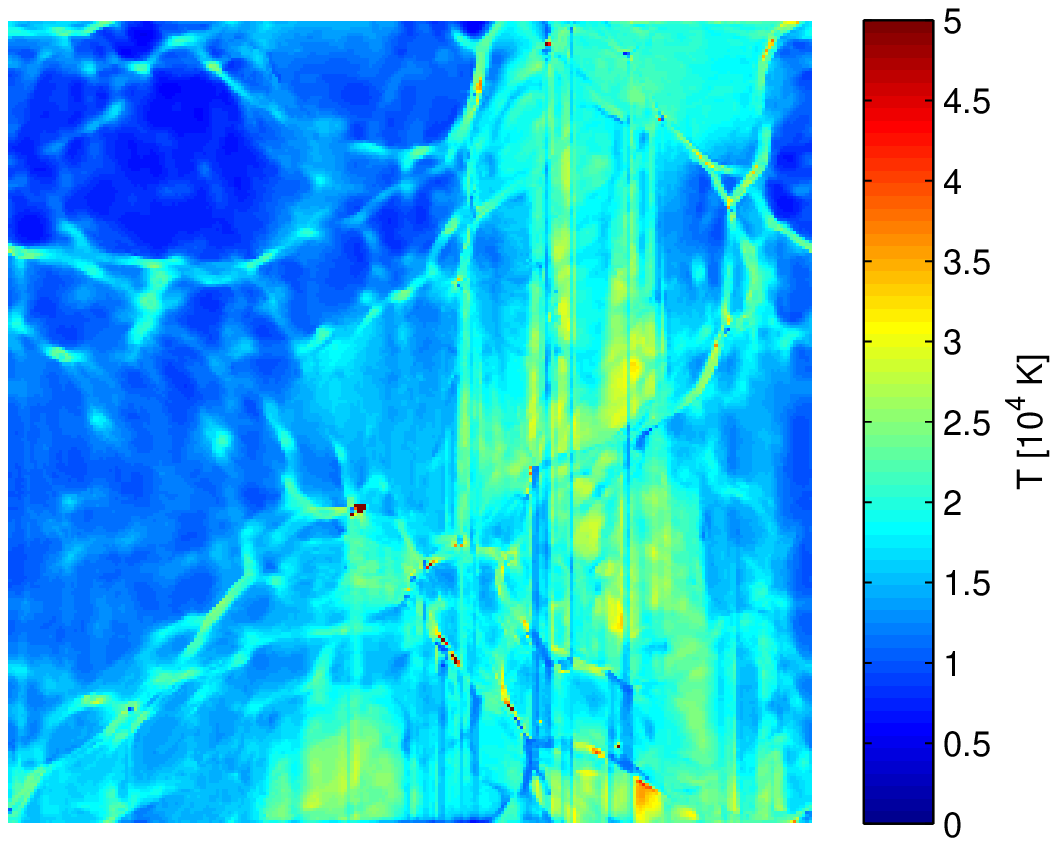}}
\caption{The post-ionization temperature for a \HeII\ ionization front
  sweeping across the box from top to bottom, at $z=3.3$ (left panel)
  and $z=2.0$ (right panel). The box side is $25h^{-1}\,$~Mpc
  (comoving).
}
\label{fig:Temp_map}
\end{figure*}

\begin{figure*}
\scalebox{0.6}{\includegraphics{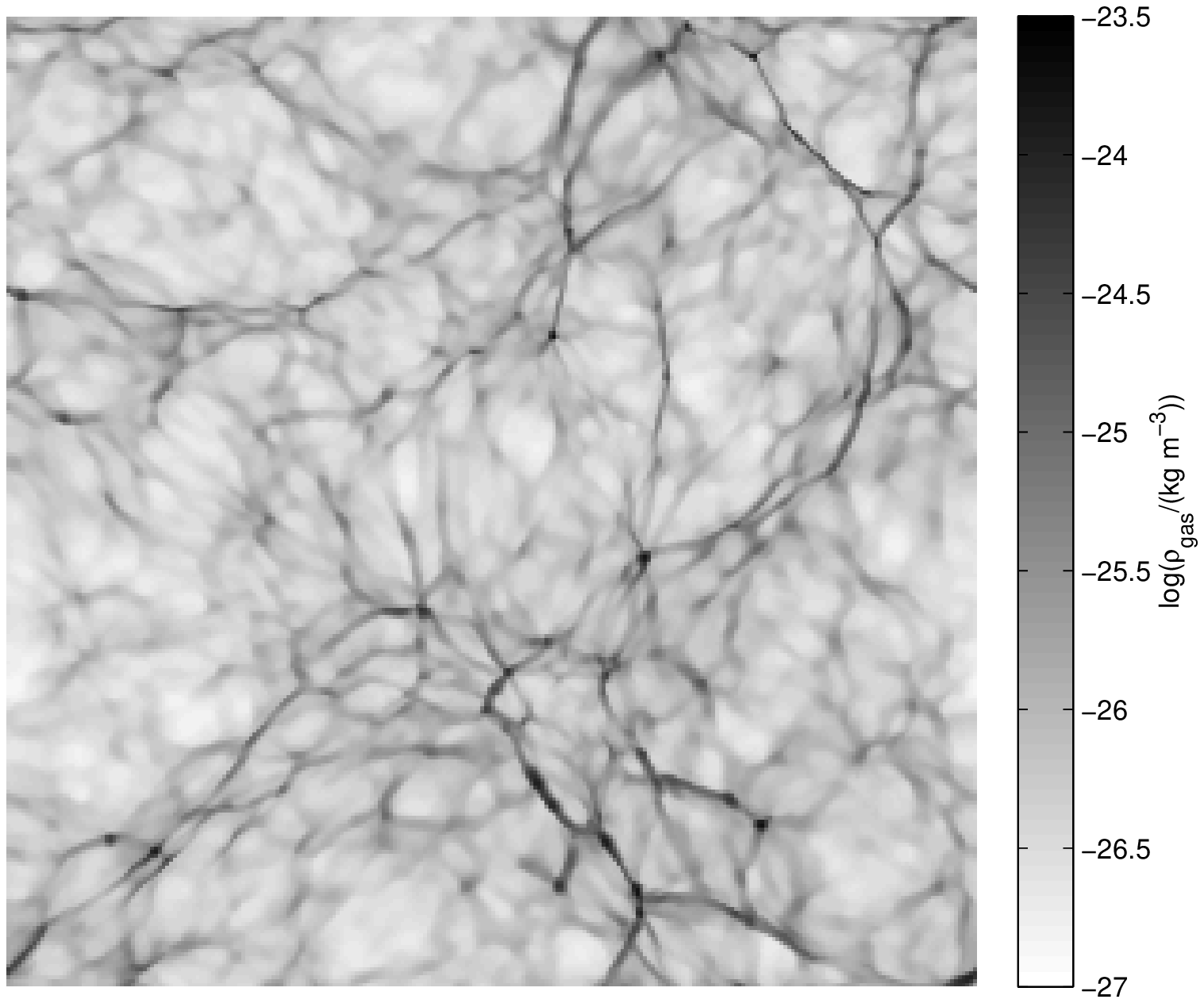}\quad\qquad\includegraphics{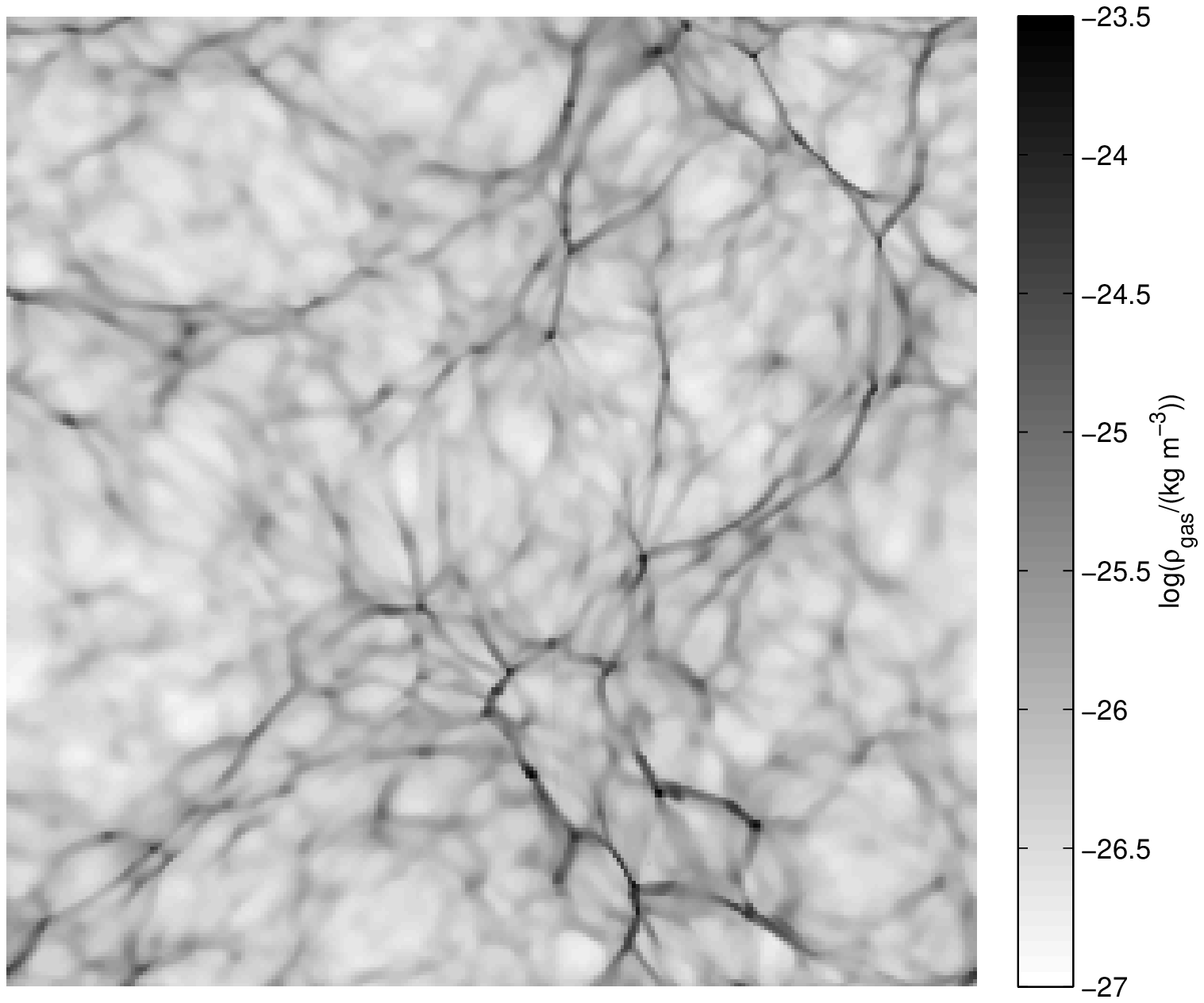}}
\caption{The gas distribution at $z=2.0$ without \HeII\ reionization
  (left panel) and with (right panel). Gas is driven out of small
  haloes by the \HeII\ reionization. The box side is $25h^{-1}\,$~Mpc
  (comoving).
}
\label{fig:Gas_density_map}
\end{figure*}

The resulting temperature structure is shown in
Fig.~\ref{fig:Temp_map} as the \HeII\ I-front passes through the box
at $z=3.3$ and at $z=2.0$, after \HeII\ ionization has completed. The
effect of shadowing produced by density inhomogeneities in the IGM are
apparent as the advance of the ionization front is delayed beyond the
densest regions. This may best be seen by comparing the temperature
map at $z=3.3$ in Fig.~\ref{fig:Temp_map} with the density map in
Fig.~\ref{fig:Gas_density_map}. (The principal density structures
change little between $z=3.3$ and $z=2.0$ in comoving coordinates.)
Large temperature boosts of $3-4\times10^4$~K occur downstream from
dense clumps, consistent with the striations of high heating rates per
ionization in Fig.~\ref{fig:epsHeII} as expected for photoionization
by a radiation field hardened by intervening optically thick clumps
with optical depths of several tens, as discussed in
Sec.~\ref{sec:photoheating}.

By $z=2.0$, most of the high temperature striations have been replaced
by gas temperatures that trace the matter distribution now that \HeII\
reionization has completed. Pronounced hot regions, however, persist,
particularly in underdense regions unable to achieve thermal
equilibrium. These lay behind dense structures while the gas was
reionized. The filtering of the radiation field by the dense
structures hardened the spectrum of the incident radiation passing
downstream to the underdense gas, as shown in Fig.~\ref{fig:epsHeII},
producing the high temperatures. This effect is especially prevalent
in the centre-right portion of the figure.

\subsection{Differential impact}
\label{subsec:diffimp}

\begin{figure}
\scalebox{0.6}{\includegraphics{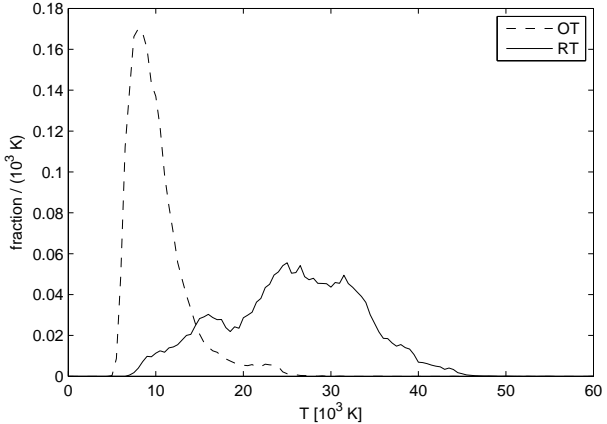}}
\caption{Distribution of temperatures at $z=3$ following reionization
  including radiative transfer (solid line) vs reionization in the
  optically thin limit (dashed line).
}
\label{fig:T_hist}
\end{figure}

\begin{figure*}
\scalebox{0.65}{\includegraphics{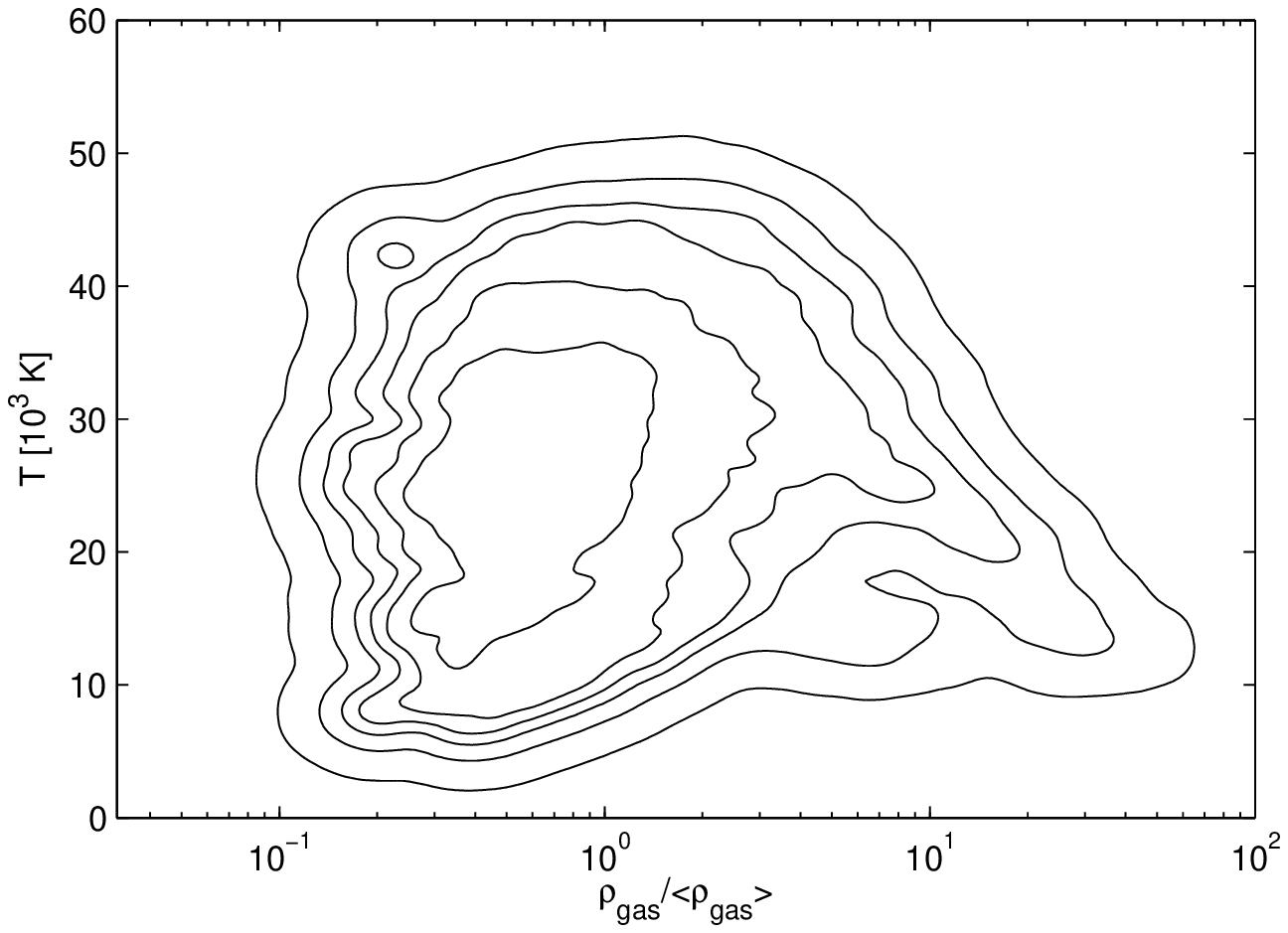}\includegraphics{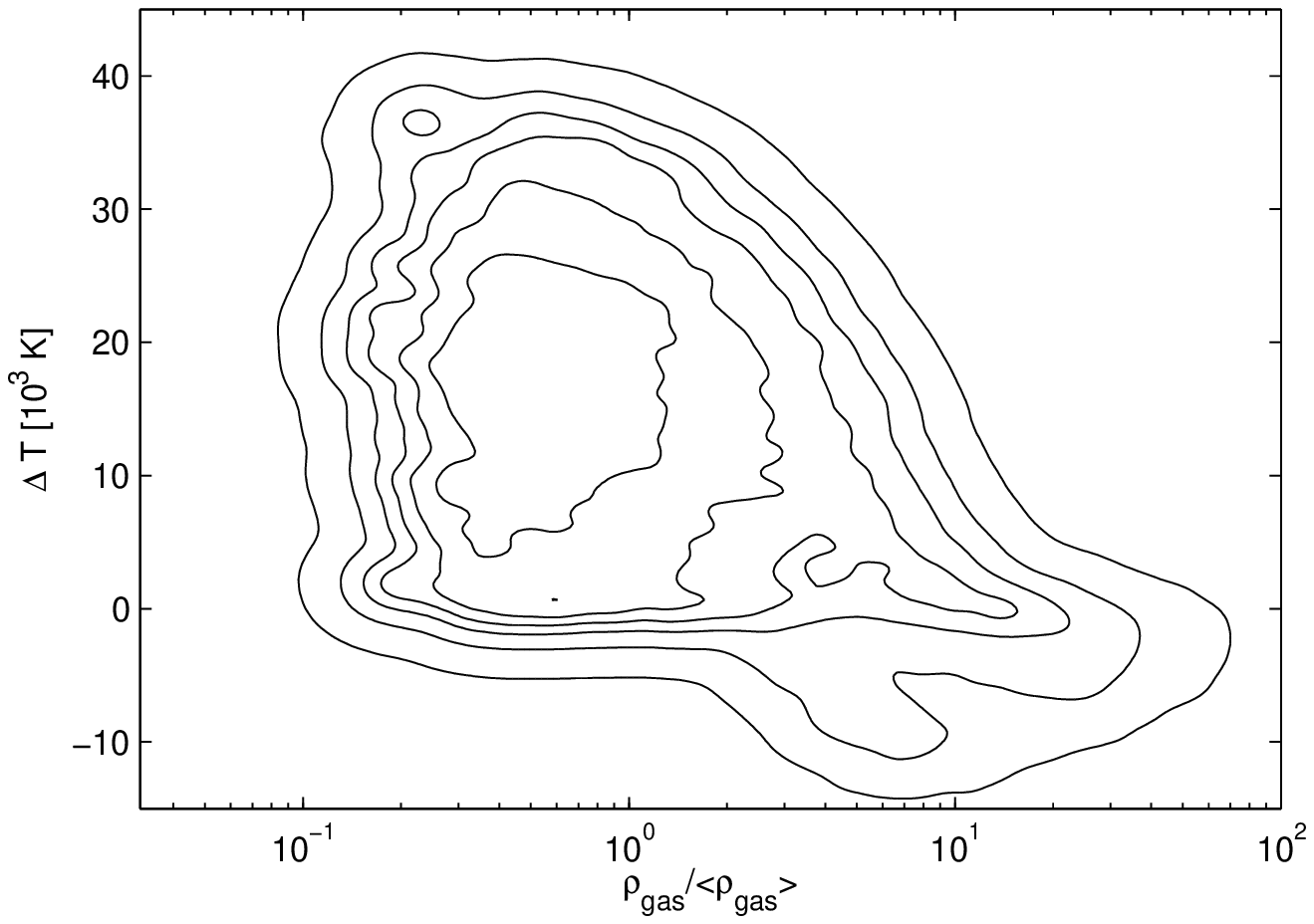}}
\caption{(Left panel) The temperature as a function of gas density at
  $z=3.0$ following \HeII\ reionization including radiative
  transfer. (Right panel) The boost in temperature as a function of
  gas density at $z=3.0$ including radiative transfer vs reionization
  in the optically thin limit. The contour levels are probability
  density levels (per $\diff
  T$-$\diff\log_{10}\rho/\langle\rho\rangle$) stepped by 0.5 dex
  starting at $10^{-7}$.
}
\label{fig:dT_rho_contours}
\end{figure*}

Including radiative transfer in the ionization of \HeII\ substantially
boosts the temperature of the gas, increasing it typically by
$15-20\times10^3$~K over the optically thin reionization limit, as
shown in Fig.~\ref{fig:T_hist}. The temperature-density relation is
shown in Fig.~\ref{fig:dT_rho_contours}. The plateau with a gas
temperature of $\sim20-30\times10^3$~K at the mean baryon density
agrees well with the estimate of \citet{Schaye00} of
$T_0\sim20-30\times10^3$~K at $z=3$, with only a weakly increasing
temperature with overdensity, as inferred from the Doppler widths
measured in high resolution spectra of the \Lya\ forest. The
temperature found in the simulation, however, is much larger than the
temperature at the mean density of $T_0\sim10-15\times10^3$ determined
from the flux curvature method applied to high resolution data
\citep{2011MNRAS.410.1096B}. The discrepancy may be due to an
inadequate simulation model for calibrating the statistic used to
estimate the gas temperature. Alternatively, it may indicate that
either \HeII\ was reionized too late in the simulation here, or that
the \HeII-ionizing source was too hard. A temperature inversion occurs
for $\rho/\langle\rho\rangle>5$, as is expected for thermal balance
between photoionization heating and atomic line and radiative
recombination cooling in dense structures. Only such high density gas
is able to achieve thermal balance; the time scale to achieve
equilibrium is too long in underdense gas \citep{Meiksin94, MR94}. A
similar trend between temperature and density is found for a pure dark
matter simulation with radiative transfer \citep{2007MNRAS.380.1369T}.

Compared with optically thin reionization, the excess temperature
boost allowing for radiative transfer occurs primarily for underdense
gas, as shown in Fig.~\ref{fig:dT_rho_contours} (right panel). Most of
the gas with $\rho/\langle\rho\rangle > 3$ experiences very little
boost at all, as it is able to quickly recover thermal balance.

\begin{figure}
\scalebox{0.85}{\includegraphics{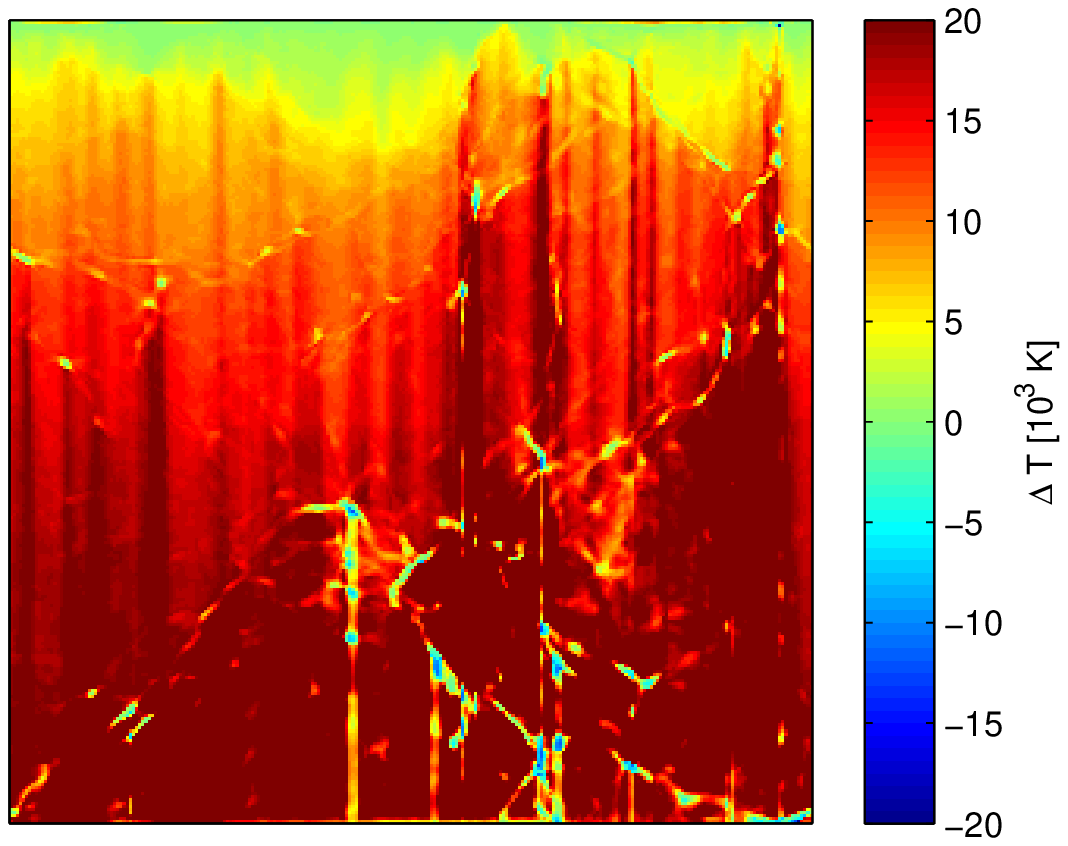}}
\caption{Map showing the change in the temperature including radiative
  transfer and without, at $z=3.0$, following \HeII\ reionization. The
  box side is $25\,h^{-1}\,{\rm Mpc}$ (comoving).
}
\label{fig:dT_map}
\end{figure}

\begin{figure*}
\scalebox{0.85}{\includegraphics{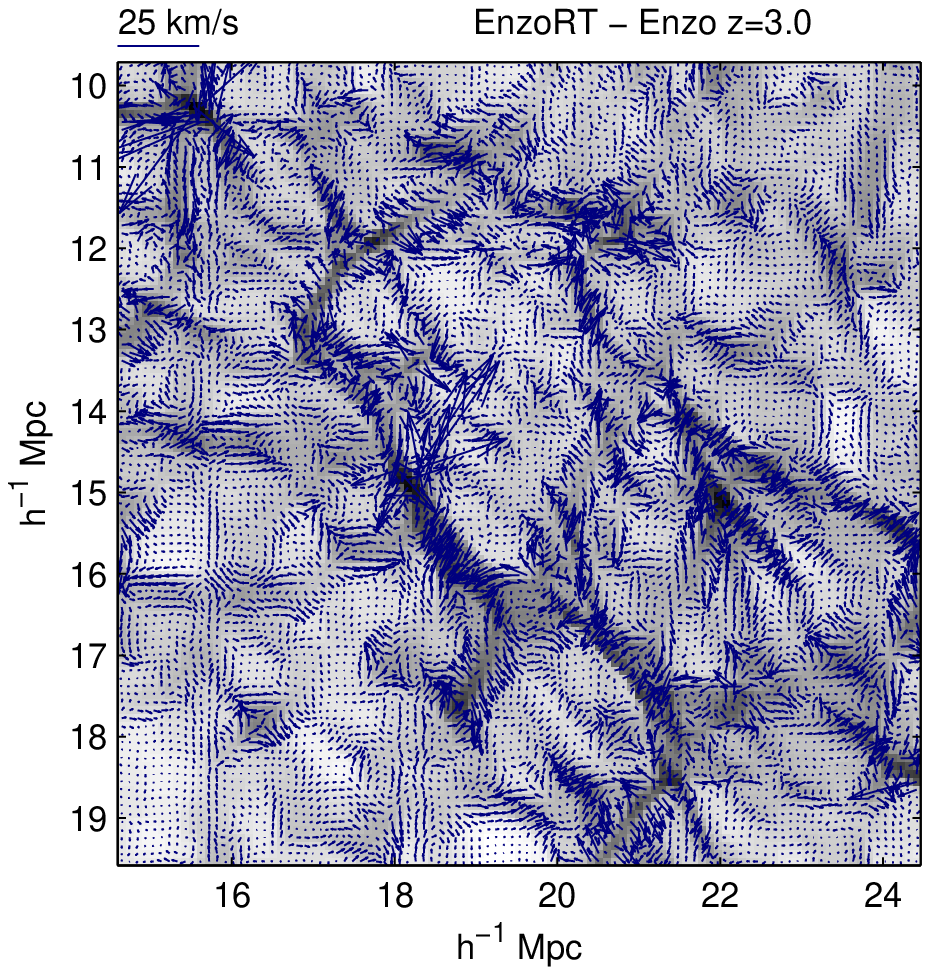}\includegraphics{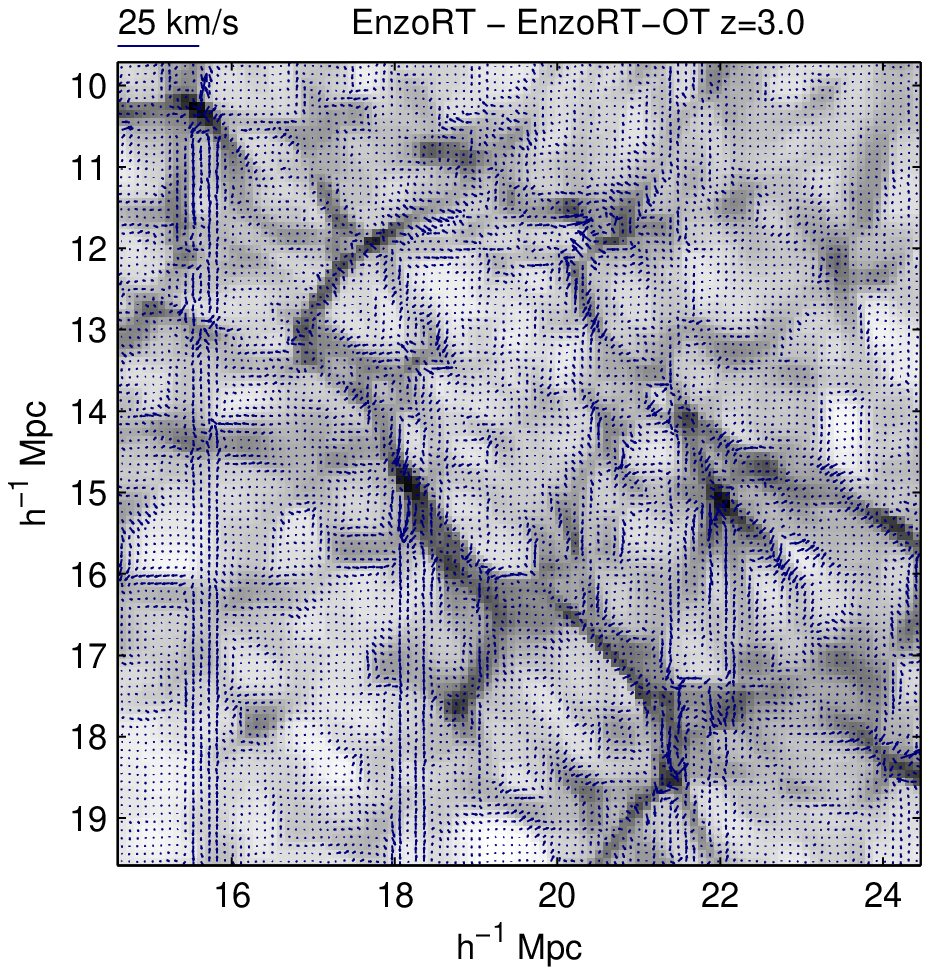}}
\caption{Differential peculiar velocity map in a kinematically complex
  region at $z=3.0$. (The indicated box length scale is comoving.)
  Left panel:\ Map showing the change in the projected peculiar
  velocity of the gas with \HeII\ reionization including radiative
  transfer from the case without \HeII\ reionization. Right panel:\
  Map showing the change in the projected peculiar velocity with
  \HeII\ reionization including radiative transfer from the case with
  \HeII\ reionization in the optically thin approximation.
}
\label{fig:pecv_map}
\end{figure*}

\begin{figure}
\scalebox{0.6}{\includegraphics{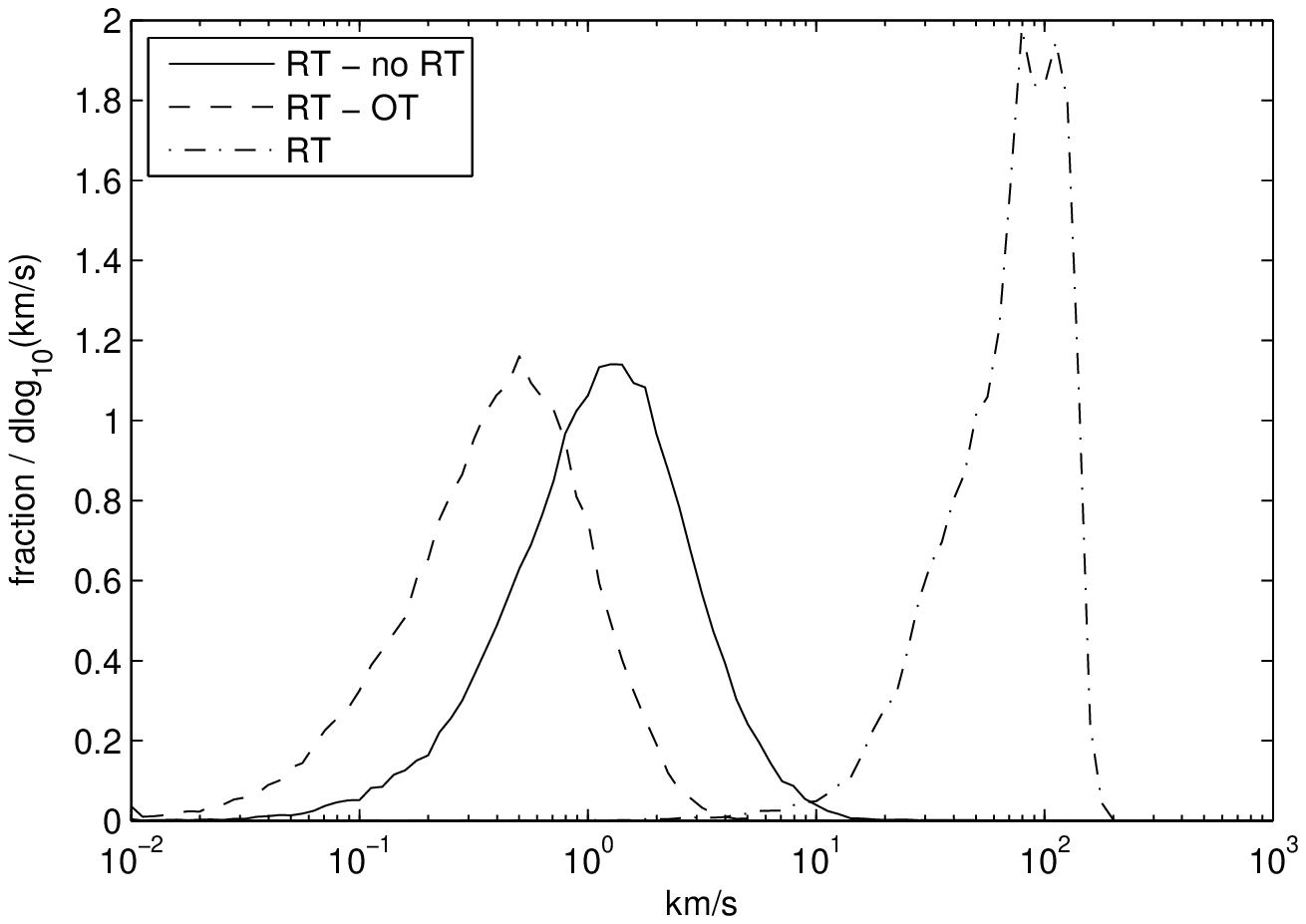}}
\caption{Distribution of peculiar velocity difference at $z=3$ for
  reionization including radiative transfer vs the case with no
  reionization (solid line) and reionization in the optically thin
  limit (dashed line). Also shown is the total peculiar velocity
  distribution for the reionization simulation including radiative
  transfer (dot-dashed line).
}
\label{fig:pecv_hist}
\end{figure}

A map of the difference in the temperatures is shown in
Fig.~\ref{fig:dT_map}. Boosts of $\Delta T>15\times10^3$~K are visible
downstream of very dense structures which act to filter and harden the
radiation field passing through them. The densest structures
themselves show little change in temperature, as the temperature
rapidly reaches thermal balance. Small overdense patches, however,
show marked temperature deficits, as much as $\Delta T <
-10\times10^3$~K. These regions account for the temperature reductions
in Fig.~\ref{fig:dT_rho_contours}. The temperature decreases arise
from complete shadowing of the \HeII-ionizing photons by an
intervening optically thick system at the \HeII\ photoelectric
edge. While for optically thin reionization, all the gas everywhere is
exposed to the photoionizing background, allowing for radiative
transfer will produce shadowed regions in the presence of sufficiently
dense clumps of gas. Only after \HeII\ reionization completes will the
regions be reheated by overlapping ionization fronts.

The increase in the peculiar velocity compared with the computation in
the optically thin approximation is shown in Fig.~\ref{fig:pecv_map}
(right panel). The differences are small, typically less than $1\kms$
as shown in Fig.~\ref{fig:pecv_hist} (dashed curve), except near the
most overdense structures, or behind overdense regions which shadow
the incident radiation field and delay full helium reionization. The
small magnitudes are in contrast to the peculiar velocity boosts
induced by reionization compared with the non-reionization case, shown
in the left panel. Reionization itself induces sizeable peculiar
velocities, typically $0.1<\Delta v < 10\kms$, throughout most of the
overdense gas, as shown in Figs.~\ref{fig:pecv_map} (left panel) and
\ref{fig:pecv_hist} (solid curve).

\begin{figure*}
\scalebox{0.85}{\includegraphics{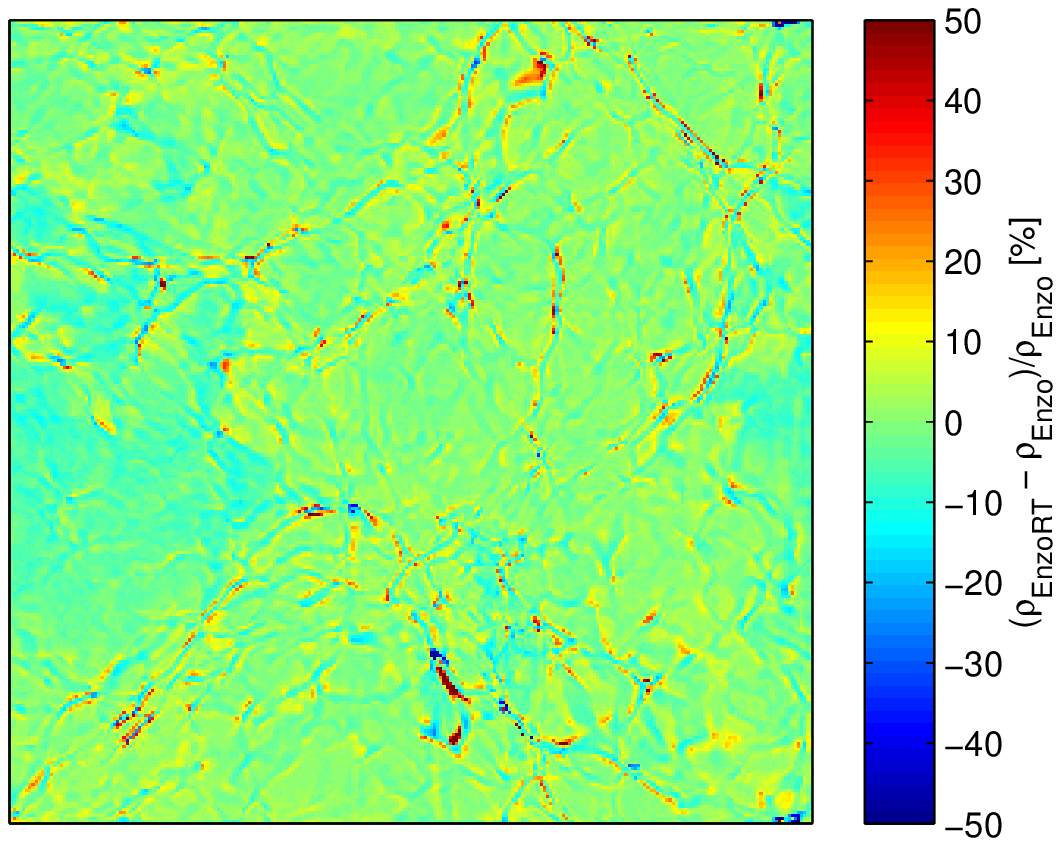}\includegraphics{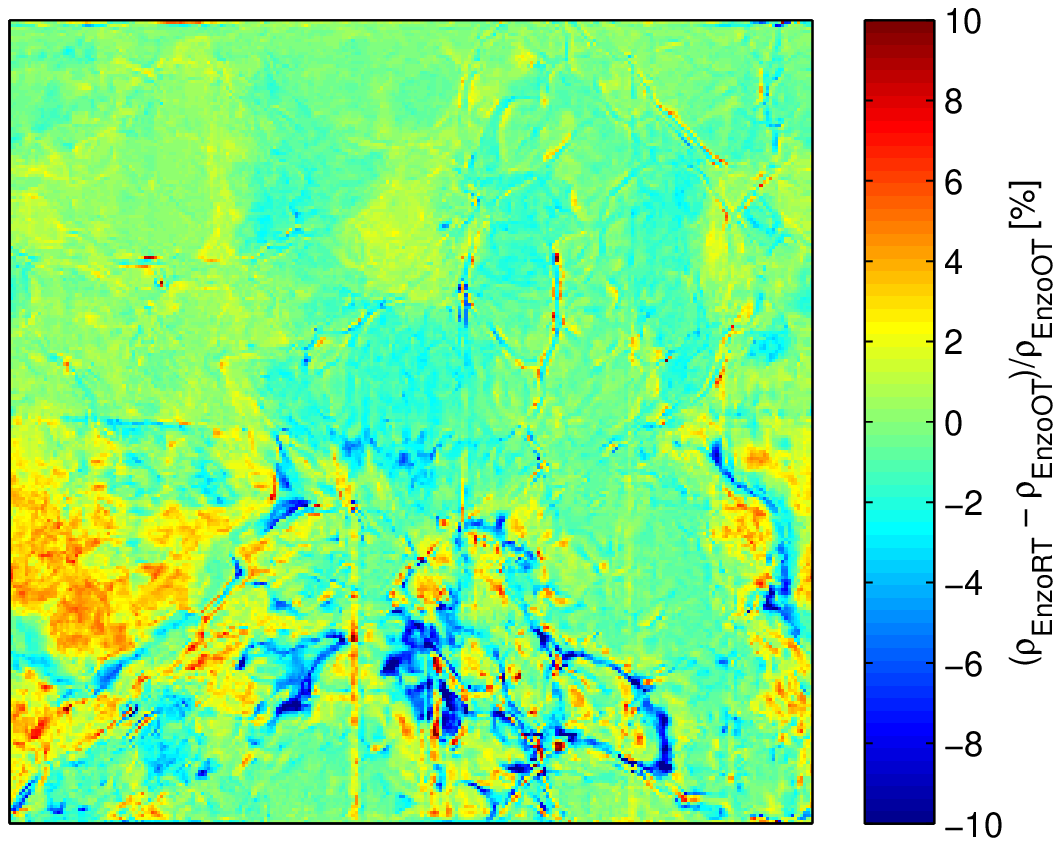}}
\caption{Maps showing the differential effect of \HeII\ reionization
  on the gas density at $z=3$. Left panel:\ Map showing the percentage
  change in the gas density with \HeII\ reionization including
  radiative transfer from the case without \HeII\ reionization. Right
  panel:\ Map showing the percentage change in the gas density with
  \HeII\ reionization including radiative transfer from the case with
  \HeII\ reionization in the optically thin approximation. (Note
  change in scale.) The box side is $25h^{-1}\,$~Mpc (comoving).
}
\label{fig:GasDensityDiff}
\end{figure*}

\begin{figure*}
\scalebox{0.85}{\includegraphics{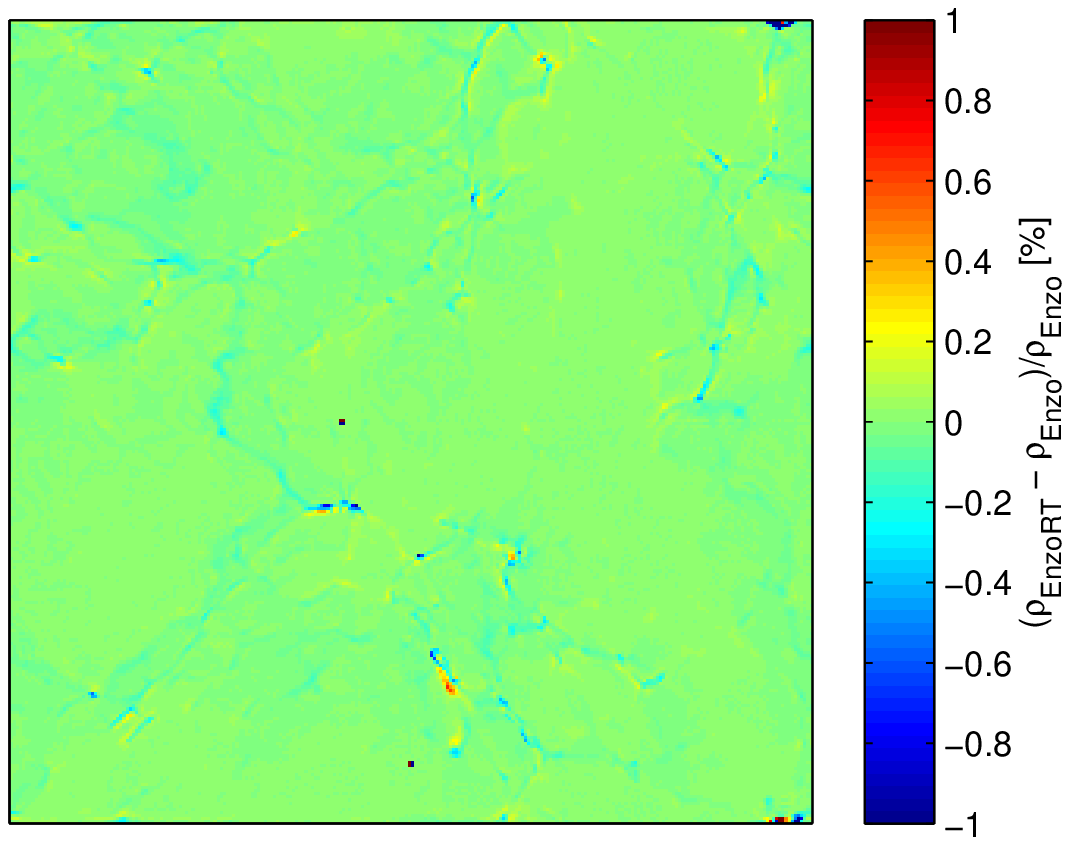}\includegraphics{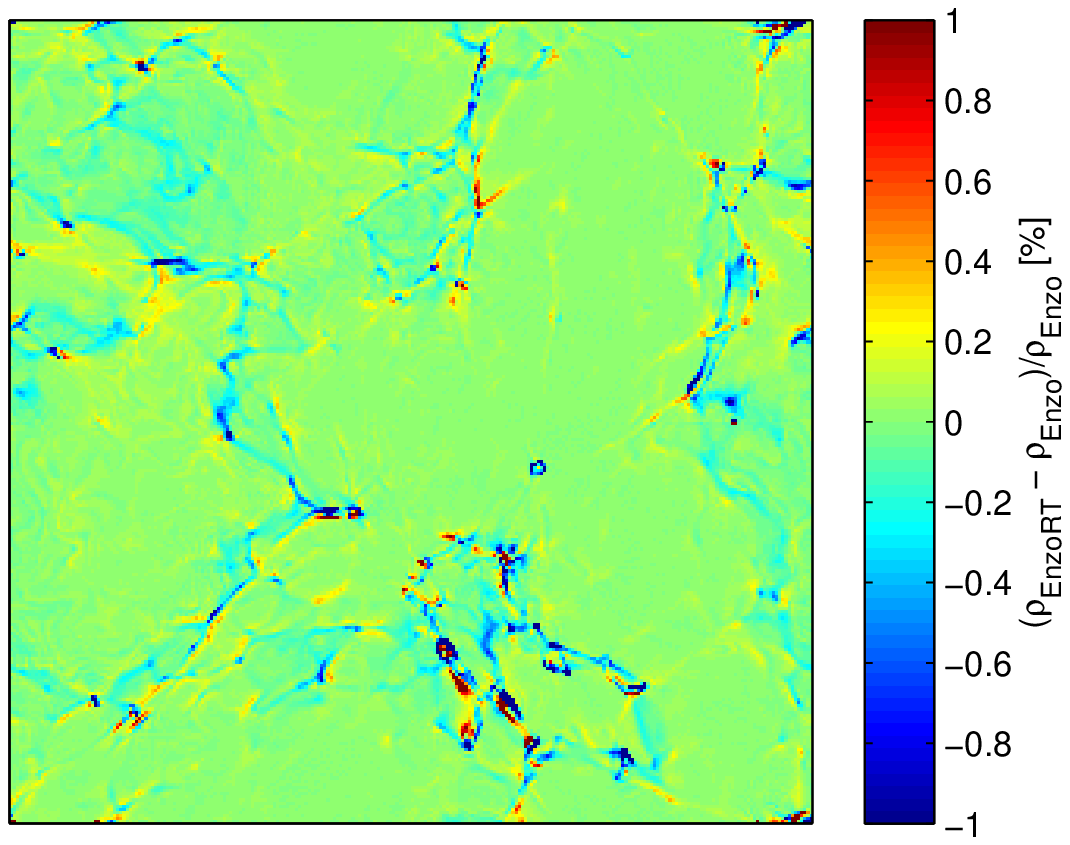}}
\caption{Maps showing the differential effect of \HeII\ reionization
  on the dark matter density at $z=3$ (left panel) and $z=2$ (right
  panel). The box side is $25h^{-1}\,$~Mpc (comoving).
}
\label{fig:DMDensityDiff_Enzo_EnzoRT}
\end{figure*}

\begin{figure*}
\scalebox{0.85}{\includegraphics{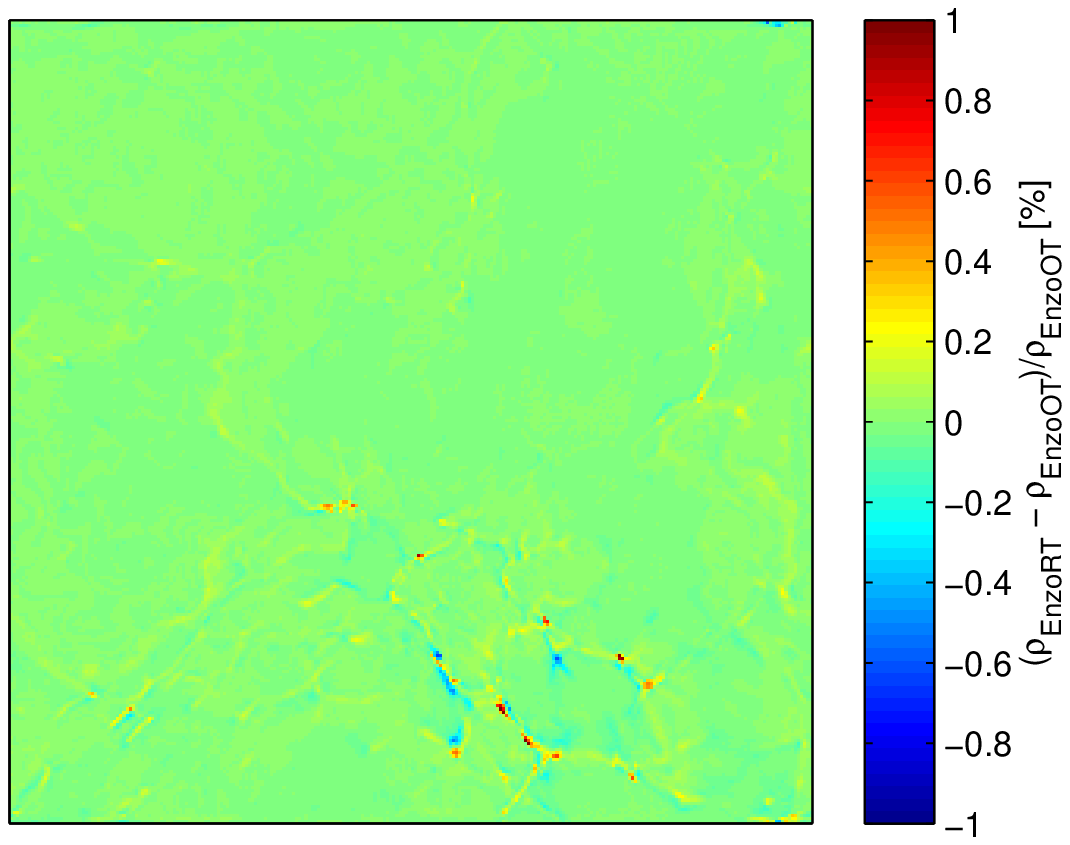}\includegraphics{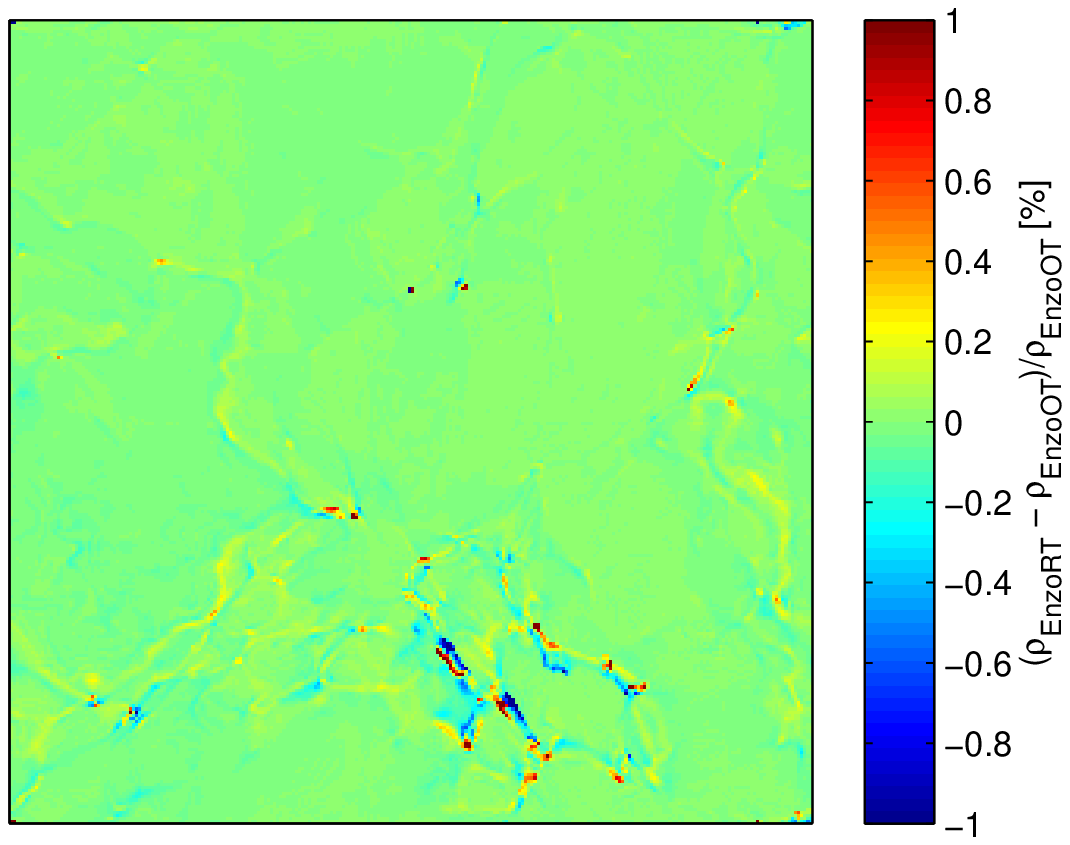}}
\caption{Maps showing the differential effect of \HeII\ reionization
  on the dark matter density allowing for full radiative transfer
  compared with \HeII\ reionization in the optically thin
  approximation. Shown at $z=3$ (left panel) and $z=2$ (right
  panel). The box side is $25h^{-1}\,$~Mpc (comoving).
}
\label{fig:DMDensityDiff_EnzoRT_EnzoOT}
\end{figure*}

A comparison of the two panels in Fig.~\ref{fig:Gas_density_map} shows
that some of the dense gaseous clumps present in the run without
\HeII\ reionization disperse when \HeII\ reionization is included,
particularly in the complex filamentary regions in the lower half of
the figure. The corresponding boosts in the outflows following \HeII\
reionization are shown in the left panel of Fig.~\ref{fig:pecv_map}.
The differential impact of \HeII\ reionization on the gas density of
the IGM is shown in Fig.~\ref{fig:GasDensityDiff} (left
panel). Substantial changes are found, with a $\sim30$ per
cent. decrease in underdense regions as the increased pressure
gradients resulting from the additional heat input drives gas out of
the shallower potential wells. Enhancements are produced as well, with
the density increasing by as much as 50 per cent. in dense regions
when helium reionization is included. These regions tend to trace the
filamentary structure of the gas distribution, suggesting they are too
hot to accrete as readily on to nearby haloes.

Allowing for \HeII\ reionization, but in the optically thin limit,
produces a similar gas density field to that with radiative transfer,
as shown in Fig.~\ref{fig:GasDensityDiff} (right panel). Large
differences, however, occur in the more complex density regions, with
the radiative transfer case generally resulting in more completely
evacuated underdense regions.

The gas motion following \HeII\ reionization also produces a change in
the dark matter distribution, although at a much smaller level. As gas
is driven out of filamentary structures, the reduced gravitational
potential will result in the dark matter readjusting to a somewhat
lower density, as shown in Fig.~\ref{fig:DMDensityDiff_Enzo_EnzoRT}.
The differences are small at $z=3$, on the order of 1 percent or
less. By $z=2$, however, the differences have grown to as high as 3
per cent., with coherent structures several comoving megaparsecs in
scale showing underdensities. As shown in
Fig.~\ref{fig:DMDensityDiff_EnzoRT_EnzoOT}, most of the differences
are accounted for in the optically thin limit, but not entirely,
especially in complex density regions such as in the lower middle
region of the figure. Large-scale simulations with multiple QSO
sources are necessary to estimate the impact on the matter power
spectrum $P(k)$, however the results here suggest reionization may
perturb $P(k)$ by a few per cent. on comoving scales $k>0.5\,h{\rm
  Mpc}^{-1}$ by $z=2$.

\section{Spectral signatures}

\begin{table}
\caption{Effective IGM optical depths for \HI\ and \HeII\ \Lya.}
\begin{tabular}{lccccc}
\hline & \multicolumn{5}{c}{$z$}\\Component & 2.0 & 2.5 & 3.0 & 3.5 & 4.0 \\
            \hline $\tau_{\rm eff}^{\rm HI}$ & 0.13 & 0.23 & 0.40 &
            0.72 & 0.89 \\
            \hline $\tau_{\rm eff}^{\rm HeII}$ & 0.45 & 2.0 & 5.0 &
            $\dots$ & $\dots$ \\ \hline
\label{tab:taueff}
\end{tabular}
\end{table}

To assess the impact of radiative transfer on the spectral signatures
of the \Lya\ forest, spectra against a ficticious background source
are drawn from the {\texttt{Enzo}} simulations with and without
radiative transfer. The spectra are generated following the procedure
described in \citet{2007MNRAS.380.1369T}. They are constructed at an
angle relative to the simulation axes in order to create a long
non-repeating spectrum that wraps around a slice through the
simulation volume. The velocity pixel widths are $5\kms$. Both \HI\
\Lya\ and \HeII\ \Lya\ spectra are made. The spectra are normalised to
a mean transmission of $\langle\exp(-\tau)\rangle=\exp(-\tau_{\rm
  eff})$ with the values given in Table~\ref{tab:taueff} for \HI\
\citep{2008ApJ...681..831F} and \HeII\ \citep{Zheng04,
  2010arXiv1008.2957S}. Absorption lines are fit to the spectra using
{\texttt{AUTOVP}} \citep{1997ApJ...477...21D}, modified to recover
from fatal errors \citep{MBM01}. At $z=3.0$, the \HeII\ \Lya\
transmission is very patchy and found too small to readily fit
absorption line features.

\subsection{\HI\ absorption signature}

\begin{figure}
  \scalebox{0.45}{\includegraphics{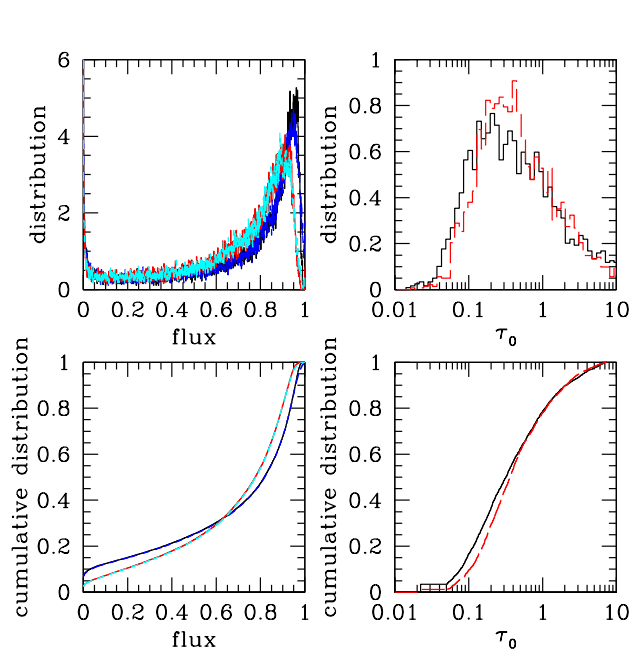}}
  \caption{(Top left panel) The pixel flux distribution function at
    $z=3$ following \HeII\ reionization for {\texttt{Enzo}} with
    radiative transfer (black solid line) and in the optically thin
    limit (red solid line). Also shown is the recovery of the flux
    distribution using only the fit absorption features for the run
    with radiative transfer (blue short-dashed line) and without (cyan
    short-dashed line). (Bottom left panel) The corresponding
    cumulative flux distribution. (Top right panel) The distribution
    of line centre optical depths for the fit lines for reionization
    including radiative transfer (black solid line) and without (red
    long-dashed line). (Bottom right panel) The corresponding
    cumulative distribution.
}
\label{fig:fluxdist_recon_EnzoRT_EnzoRT-OT_z3}
\end{figure}

The pixel flux distribution at $z=3.0$ is shown following \HeII\
reionization both with and without radiative transfer in
Fig.~\ref{fig:fluxdist_recon_EnzoRT_EnzoRT-OT_z3} (left
panels). Allowing for radiative transfer produces a smaller fraction
of low optical depth pixels ($0.1<\tau<0.6$), and a larger fraction of
very low values ($\tau<0.1$), compared with the optically thin
approximation (right panels).

Also shown are the flux distributions recovered by fitting absorption
lines to the spectra, and regenerating the flux distribution from the
absorption lines alone. These are found to accurately recover the
original flux distribution, demonstrating the absence of residual
absorption that may not be accounted for by absorption lines.

The distribution of line centre optical depths is provided in the
right panels of Fig.~\ref{fig:fluxdist_recon_EnzoRT_EnzoRT-OT_z3},
showing a higher fraction of lines with line centre optical depth
$\tau_0<0.1$ for the case with radiative transfer compared with the
optically thin limit. The higher proportion of very low fluctuations
when radiative transfer is included may appear counter-intuitive since
the temperatures, and so thermal pressures, are greater. The increased
Jeans length may have been expected to smooth away small fluctuations
in the spectra. Consideration of the dependence of the optical depth
on the thermal width provides an explanation. Since the line centre
optical depth varies as $\tau_0\propto N_{\rm HI}/ b$, for a given
column density, an increased Doppler parameter due to heating the gas
to a higher temperature reduces the line centre optical depth. The
increase in temperature will also decrease the recombination rate at a
given gas density, and so reduce the \HI\ column density.

\begin{figure}
  \scalebox{0.43}{\includegraphics{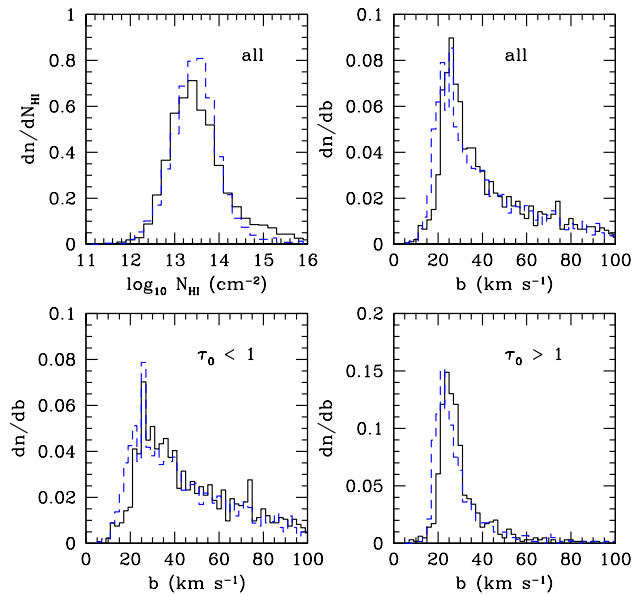}}
  \caption{(Top left panel) The \HI\ column density distribution at
    $z=3$ following \HeII\ reionization for {\texttt{Enzo}} with
    radiative transfer (solid line; black) and in the optically thin
    limit (dashed line; blue). (Top right panel) The corresponding
    Doppler parameter distribution. (Lower left panel) The Doppler
    parameter distribution for absorption lines with line centre
    optical depth $\tau_0<1$. (Lower right panel) The Doppler
    parameter distribution for absorption lines with $\tau_0>1$.
}
\label{fig:bdist_EnzoRT_EnzoRT-OT}
\end{figure}

The column density and Doppler parameter distributions for the
absorption line fits to the spectra are shown in
Fig.~\ref{fig:bdist_EnzoRT_EnzoRT-OT}. The column density
distributions for the simulations with and without radiative transfer
are very similar, with the distribution for the RT case somewhat
broader. By contrast, the Doppler parameter distribution for the RT
case shows a higher cutoff at the low end compared with the optically
thin case. Dividing the lines into those optically thin and optically
thick at line centre reveals that a large contribution to the offset
originates from the optically thin lines. The median Doppler parameter
of the optically thin absorption lines for the RT case is $b_{\rm
  med}\simeq45\kms$, while for the optically thin case, $b_{\rm
  med}\simeq41\kms$. For the optically thick absorbers, $b_{\rm
  med}\simeq26\kms$ for the RT case, while for the optically thin
reionization case $b_{\rm med}\simeq24\kms$. Since the optically thin
absorbers arise in moderate to low overdensity structures, the gas in
the optically thin systems is too rarefied to establish thermal
balance between radiative heating and atomic cooling. As a
consequence, the systems are overheated \citep{MR94, Meiksin94,
  MBM01}.

A similar trend was found for the optically thin reionization models
of \citet{MBM01} compared with the data:\ the median Doppler parameter
of the measured optically thin lines were broader than the lines drawn
from the best-fitting $\Lambda$CDM model simulation. The median
Doppler parameter for lines with $\tau_0<1$ measured in the spectrum
of Q1937--1009 \citep{1997AJ....114.1330B}, over the redshift range
$3.1<z<3.7$ was found to be $b_{\rm med}\simeq29\kms$, while the
best-fitting model gave $b_{\rm med}\simeq20\kms$. The measured median
value for the systems with $\tau_0<1$ is substantially smaller than
found here for the RT simulation, suggesting the \HeII\ reionization
was too recent in the simulation (cf. \citet{2007MNRAS.380.1369T,
  2010MNRAS.401...77M}), or that the input spectrum too hard on
average. By contrast, the data for the $\tau_0>1$ systems gave $b_{\rm
  med}\simeq26\kms$, in agreement with the RT simulation result found
here. A fully fair comparison with the observations, however, would
require replicating the redshift range, resolution and noise
properties of the measured spectra, as well as varying the
luminosities, spectral shapes and timings of the sources. Such a
comparison would require a large suite of simulations, which is not
the purpose of this paper.

\begin{figure}
  \scalebox{0.43}{\includegraphics{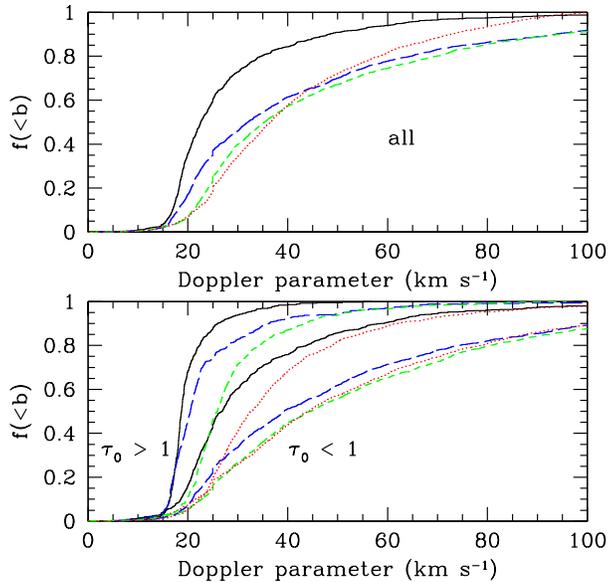}}
  \caption{(Top panel) The \HI\ Doppler parameter cumulative
    distribution for {\texttt{Enzo}} with radiative transfer, shown at
    $z=2$ (solid line; black), 2.5 (long-dashed line; blue), 3.0
    (short-dashed line; green) and 3.5 (dotted line; red). (Bottom
    panel) As in the top panel, but with the lines divided between
    those with line centre optical depth $\tau_0<1$ (rightmost curves)
    and those with $\tau_0>1$ (leftmost curves).
}
\label{fig:bcumd_EnzoRT}
\end{figure}

The evolution of the Doppler parameter cumulative distribution is
shown in Fig.~\ref{fig:bcumd_EnzoRT}. The distributions show little
evolution over $2.5<z<3.5$, but a marked decrease in the Doppler
parameters by $z=2.0$ as the IGM adiabatically cools following \HeII\
reionization. Dividing the absorption lines into those optically thick
and thin at line centre (bottom panel), reveals that the optically
thick systems undergo continual evolution towards smaller values with
cosmic time, with the median Doppler parameter evolving from $b_{\rm
  med}\simeq33\kms$ at $z=3.5$ to $19\kms$ by $z=2.0$ as the gas
establishes thermal balance following \HeII\ reionization. By
contrast, the systems optically thin at line centre show almost no
evolution between $z=3.5$ and 3.0, with heavily broadened lines
following \HeII\ reionization having a median Doppler parameter
$b_{\rm med}\simeq44\kms$, diminishing slightly to $39\kms$ at
$z=2.5$, and then substantially to $27\kms$ by $z=2.0$.

\begin{figure}
  \scalebox{0.43}{\includegraphics{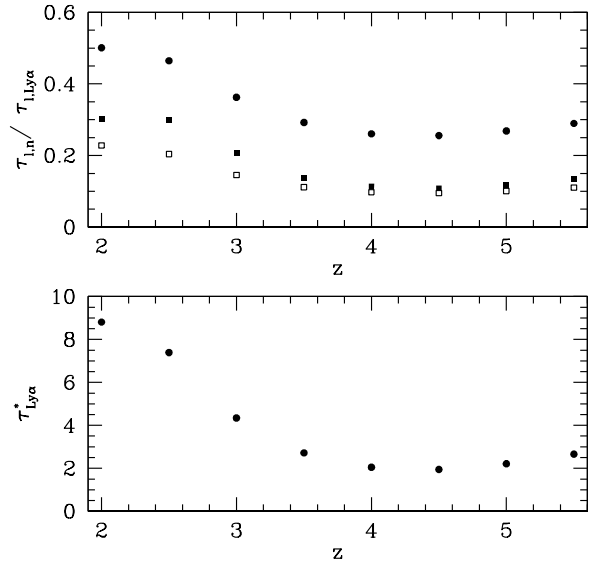}}
  \caption{(Top panel) The ratios of the effective optical depths
    $\tau_{l, {\rm Ly\beta}}/\tau_{l, {\rm Ly\alpha}}$ (solid points)
    and $\tau_{l, {\rm Ly\gamma}}/\tau_{l, {\rm Ly\alpha}}$ (solid
    squares) for \HI\ obtained from the {\texttt{Enzo}} simulation
    with radiative transfer, as a function of redshift $z$. Also shown
    is the predicted ratio $\tau_{l, {\rm Ly\gamma}}/\tau_{l, {\rm
        Ly\alpha}}$ (open squares) based on a line-blanketing model
    (see text). (Bottom panel) The characteristic \HI\ $\tau^*_{\rm
      Ly\alpha}$ optical depth inferred from the values of $\tau_{l,
      {\rm Ly\beta}}/\tau_{l, {\rm Ly\alpha}}$ from the
    {\texttt{Enzo}} simulation with radiative transfer.
}
\label{fig:tauHILynpLya}
\end{figure}

In addition to the scattering of \Lya\ photons, the higher order Lyman
resonance line photons will be scattered as well. Identifying
individual absorption features in spectra is difficult except for the
strongest lines, especially at high redshifts. The effective optical
depths resulting from line blanketing, however, may be inferred. The
observed effective optical depth at redshift $z$ due to systems with
restframe equivalent widths $w$ (in wavelength units) is given by
\begin{equation}
\tau_{l, {\rm Ly}n} =
\frac{1+z}{\lambda_{1n}}\int\,\diff w \frac{\partial^2{\sl N}}{\partial
  z \partial w}w,
\label{eq:taulLyn}
\end{equation}
where $\lambda_{1n}$ is the wavelength of the Lyman resonance line
transition to the n$^{\rm th}$ principal quantum level, and where the
equivalent width is related to the line centre optical depth
$\tau_{0,1n}$ by $w=2(b/c)\lambda_{1n}F(\tau_{0,1n})$, where $F(\tau)$
is a function dependent on the Voigt profile
\citep{2009RvMP...81.1405M}. If the line distribution is separable
into the form $\partial^2{\sl N}/\partial z \partial
w=(dN/dz)f(w/w_{1n}^*)$, where the equivalent width distribution
$f(w/w_{1n}^*)$ may be characterised by a single parameter $w_{1n}^*$,
such as the scale height for an exponential distribution
$f(w/w_{1n}^*)=\exp(-w/w_{1n}^*)$, then the optical depths for
transitions to levels $n$ and $n^\prime$ are related by
\begin{equation}
\frac{\tau_{l,n^\prime}}{\tau_{l,n}} =
\frac{F\left(\frac{\lambda_{1n^\prime}f_{1n^\prime}}{\lambda_{1n}f_{1n}}
    \tau_{0,1n}^*\right)}{F(\tau_{0,1n}^*)},
\label{eq:tauLynpLyn}
\end{equation}
where $\tau_{0,1n}^*$ is given by
$w_{1n}^*=2(b/c)\lambda_{1n}F(\tau_{0,1n}^*)$.

The ratios $\tau_{l,{\rm Ly\beta}}/ \tau_{l,{\rm Ly\alpha}}$ and
$\tau_{l,{\rm Ly\gamma}}/ \tau_{l,{\rm Ly\alpha}}$ from the
{\texttt{Enzo}} simulation with RT are shown in the upper panel of
Fig.~\ref{fig:tauHILynpLya}. A rise in both ratios occurs at redshifts
$z<4$ once \HeII\ reionization begins. The inferred values of the
characteristic \Lya\ optical depth $\tau^*_{\rm Ly\alpha}$ are shown
in the bottom panel. The corresponding predicted values of
$\tau_{l,{\rm Ly\gamma}}/ \tau_{l,{\rm Ly\alpha}}$ (upper panel) lie
close to, but somewhat below, the values measured from the simulation,
suggesting the single parameter separable model for the line
distribution provides a good description of the line blanketing, but
becomes less accurate at large values of $\tau^*_{\rm Ly\alpha}$.

\subsection{\HeII\ absorption signature}

\begin{figure}
  \scalebox{0.43}{\includegraphics{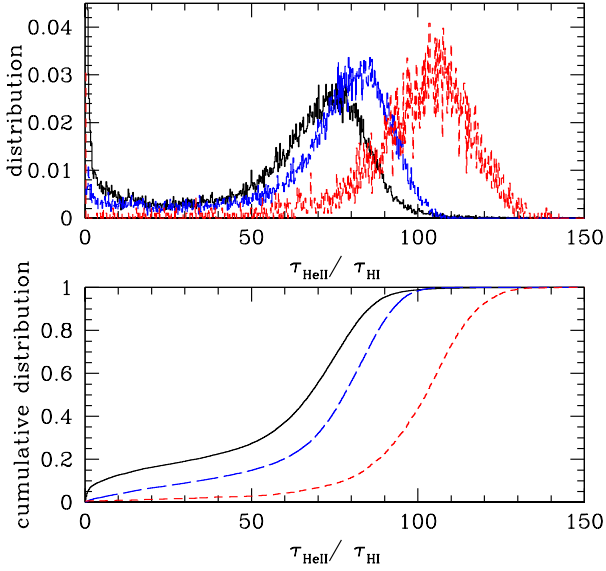}}
  \caption{(Top panel) The distributions of the ratio of the pixel
    optical depths $\tau_\HeII/\tau_\HI$ in the spectra following
    \HeII\ reionization for {\texttt{Enzo}} with radiative transfer at
    $z=2.0$ (solid line; black), 2.5 (long-dashed line; blue) and 3.0
    (short-dashed line; red). (Bottom panel) The corresponding
    cumulative distributions.
}
\label{fig:fluxdist_HeH_EnzoRT}
\end{figure}

For a homogeneous radiation field, the ratio of the \HeII\ to \HI\
optical depths per pixel is
\begin{equation}
  \frac{\tau_{\rm HeII}}{\tau_{\rm HI}} = \frac{1}{4}\frac{N_{\rm HeII}}{N_{\rm
      HI}}\frac{b_{\rm HI}}{b_{\rm HeII}},
\label{eq:tauHeIItauHI}
\end{equation}
where $N_{\rm HI}$ and $N_{\rm HeII}$ are the \HI\ and \HeII\ column
densities, respectively, and $b_{\rm HI}$ and $b_{\rm HeII}$ the
corresponding Doppler parameters. For pure thermal broadening, $b_{\rm
  HeII}=b_{\rm HI}/2$, while for velocity-broadened lines, $b_{\rm
  HeII}=b_{\rm HI}$. For metagalactic \HI\ and \HeII\ photoionization
rates $\Gamma_{\rm HI}$ and $\Gamma_{\rm HeII}$, the column density
ratio, for a hydrogen to helium number density ratio of 12.9, is
\begin{eqnarray}
\frac{N_{\rm HeII}}{N_{\rm HI}}&=&\frac{\Gamma_{\rm HI}}{\Gamma_{\rm
    HeII}}\frac{n_{\rm He}}{n_{\rm H}}\frac{\alpha_{\rm
    HeIII}}{\alpha_{\rm HI}}\nonumber\\
&\simeq&0.31\frac{\Gamma_{\rm HI}}{\Gamma_{\rm
    HeII}}\frac{7.107-(1/2)\log T+0.00547T^{1/3}}{6.414-(1/2)\log T+0.00868T^{1/3}}\nonumber\\
&\simeq&0.42\frac{\Gamma_{\rm HI}}{\Gamma_{\rm HeII}},
\label{eq:NHeIINHI}
\end{eqnarray}
where the last form is accurate to 5 per cent. for
$10^4<T<5\times10^4$~K \citep{MM94, 2009RvMP...81.1405M}. The optical
depth ratio is thus a measure of the fluctuations in the ratio of the
radiation field as well as the velocity-broadening of the absorption
lines.

The distribution of the optical depth ratio per pixel is shown in
Fig.~\ref{fig:fluxdist_HeH_EnzoRT} for $2<z<3$. The distributions show
a wide range of values, with the peak declining from 105 at $z=3$ to
75 at $z=2$. The FWHM of the distributions is about 30, changing
little with redshift. The variations arise entirely from the radiative
transfer of the incident radiation through an inhomogeneous medium
following helium reionization. Allowing for the Poisson fluctuations
from the finite number of sources within an attenuation volume for
\HeII-ionizing photons will further enhance the fluctuations
\citep{1998AJ....115.2206F, BHVC06, 2009RvMP...81.1405M,
  2010ApJ...714..355F}.

\begin{figure}
  \scalebox{0.43}{\includegraphics{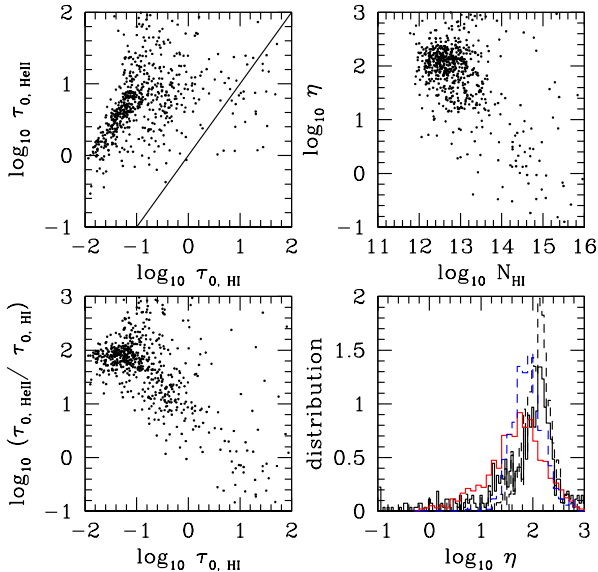}}
  \caption{Comparison of {\texttt{Enzo}} RT simulation results for
    \HI\ and \HeII\ line centre optical depths and column densities
    for fit absorption featres at $z=2.5$. (Top left panel) Comparison
    of line centre optical depths for \HI\ and \HeII\ absorption
    features. The solid line shows the relation $\tau_{0, {\rm HeII}}
    = \tau_{0, {\rm HI}}$. (Bottom left panel) \HeII\ to \HI\ line
    centre optical depth ratio {\it vs} \HI\ line centre optical
    depth. (Top right panel) \HeII\ to \HI\ column density ratio
    $\eta$ {\it vs} \HI\ column density. (Bottom right panel)
    Distribution functions of $\eta$ for all lines (solid black
    histogram), and lines restricted to $0.01<\tau_{0, {\rm HI}}<0.1$
    (dashed black histogram). Also shown are the distribution
    functions for HS~1700$+$6416 \citep{2006A&A...455...91F} for
    absorption features in the redshift range $2.30\lsim z\lsim 2.75$
    for all the lines (solid red histogram) and for lines restricted
    to $0.01<\tau_{0, {\rm HI}}<0.1$ (blue dashed histogram). All
    distribution functions are normalised to unit area.
}
\label{fig:tau0HI_tau0HeII_z2p5}
\end{figure}

\begin{figure}
  \scalebox{0.43}{\includegraphics{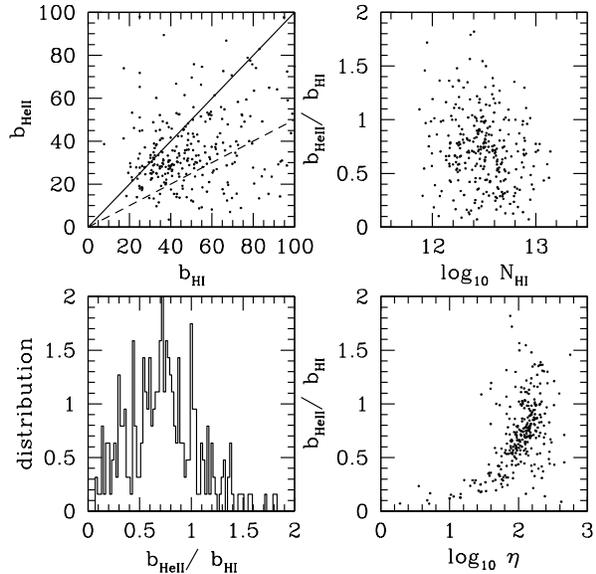}}
  \caption{Comparison of {\texttt{Enzo}} RT simulation results for
    \HI\ and \HeII\ Doppler parameters for fit absorption featres at
    $z=2.5$ with $0.01<\tau_{0, {\rm HI}}<0.1$. (Top left panel)
    Comparison of $b_{\rm HI}$ and $b_{\rm HeII}$ absorption
    features. The solid line shows the relation for pure velocity
    broadening; the dashed line shows the relation for pure thermal
    broadening. (Bottom left panel) The distribution function of
    $b_{\rm HeII}/ b_{\rm HI}$. (Top right panel) Doppler parameter
    ratio {\it vs} \HI\ column density. (Bottom right panel) Doppler
    parameter ratio {\it vs} \HeII\ to \HI\ column density ratio
    $\eta$.
}
\label{fig:bHI_bHeII_z2p5}
\end{figure}

\begin{figure}
  \scalebox{0.43}{\includegraphics{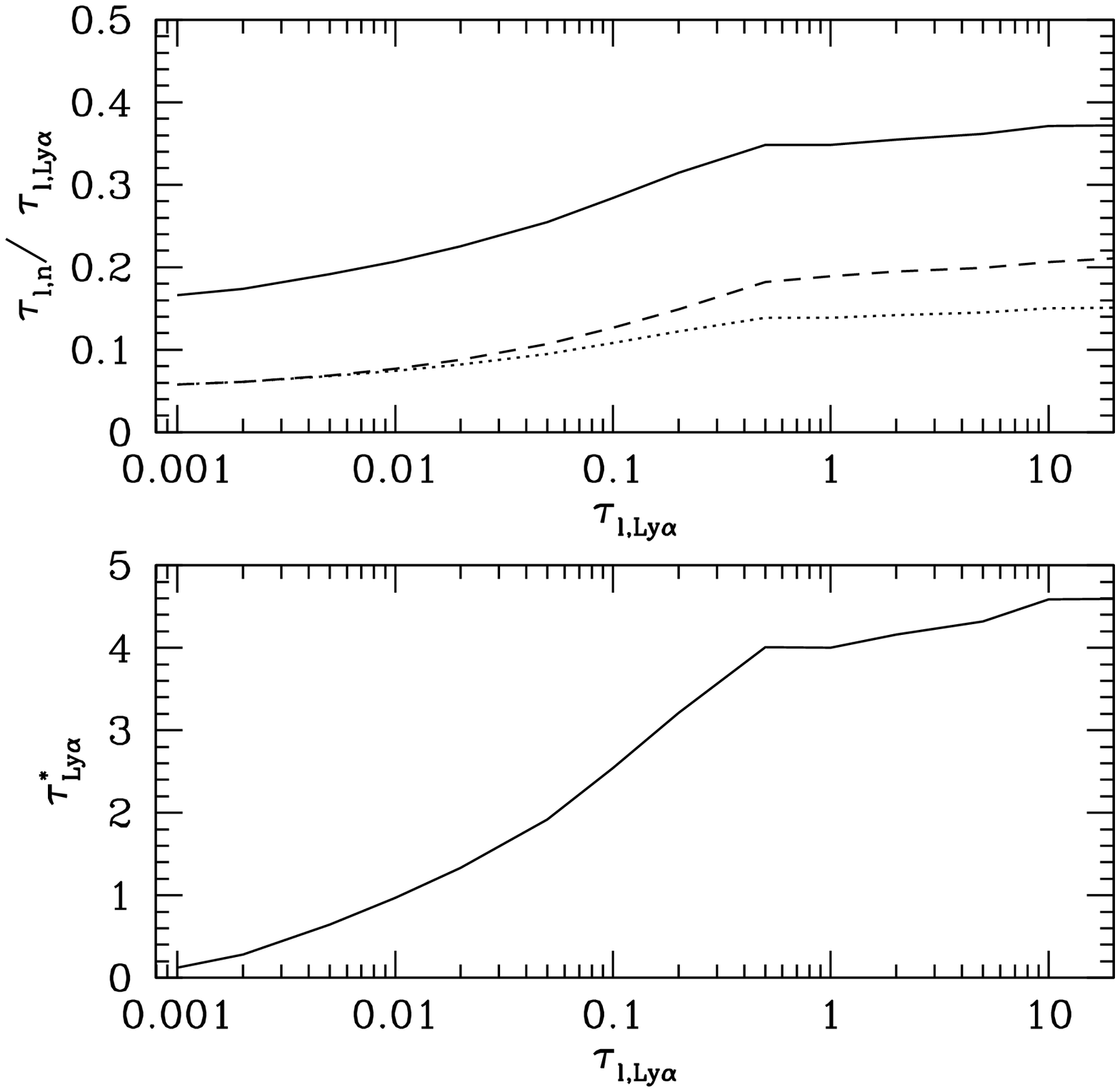}}
  \caption{(Top panel) The ratios of the effective optical depths
    $\tau_{l, {\rm Ly\beta}}/\tau_{l, {\rm Ly\alpha}}$ (solid line)
    and $\tau_{l, {\rm Ly\gamma}}/\tau_{l, {\rm Ly\alpha}}$ (dashed
    line) for \HeII\ obtained from the {\texttt{Enzo}} simulation with
    radiative transfer, as a function of $\tau_{l, {\rm
        Ly\alpha}}$. (The result is insensitive to redshift after
    reionization completes at $z\le3$.) Also shown is the predicted
    ratio $\tau_{l, {\rm Ly\gamma}}/\tau_{l, {\rm Ly\alpha}}$ (dotted
    line) based on a line-blanketing model (see text). (Bottom panel)
    The characteristic \HeII\ $\tau_{\rm Ly\alpha}^*$ optical depth
    inferred from the values of $\tau_{l, {\rm Ly\beta}}/\tau_{l, {\rm
        Ly\alpha}}$ from the {\texttt{Enzo}} simulation with radiative
    transfer.
}
\label{fig:tauHeIILynpLya}
\end{figure}

Absorption lines were fit in the \HeII\ spectrum at $z=2.5$, and
matched to the \HI\ features, requiring a line-centre velocity
difference smaller than $10\kms$ for a successful match. The relation
between the matching line centre optical depths is shown in
Fig.~\ref{fig:tau0HI_tau0HeII_z2p5}. While the deblending of
absorption features will introduce scatter into the relation, a clear
correlation is apparent in the upper left panel corresponding to
$\tau_{0, {\rm HeII}}/ \tau_{0, {\rm HI}}\lsim100$. The ratio is found
to decrease towards increasing $\tau_{0, {\rm HI}}$ (lower left
panel). This arises as the \HI\ features become saturated. A similar
trend was found for simulated spectra by \citet{2006A&A...455...91F}
for absorption lines in the redshift range $2.30\lsim z\lsim 2.75$,
who argued statistical comparisons must be restricted to systems with
$\tau_{0, {\rm HI}}<0.1$.

The distribution of column density ratios $\eta=N_{\rm HeII}/ N_{\rm
  HI}$ in Fig.~\ref{fig:tau0HI_tau0HeII_z2p5} shows a similar
decreasing trend towards increasing $N_{\rm HI}$ (top right panel). A
broad distribution of $\eta$ values for all the systems is found
(lower right panel). Restricting the systems to those with
$0.01<\tau_{0, {\rm HI}}<0.1$, however, results in a far sharper
distribution centred at $\eta\simeq130$. The distributions are very
similar to those measured \citep{2006A&A...455...91F}, with the
distribution for systems restricted to $0.01<\tau_{0, {\rm HI}}<0.1$
peaking at $\eta\lsim100$ and somewhat broader than the distribution
found from the {\texttt{Enzo}} RT simulation. The peak in the
distribution from the simulation is fixed by the normalisations
imposed on the effective \HI\ and \HeII\ optical depths. The agreement
in the widths of the distribution, however, is a product of radiative
transfer within the simulation. While the measured mean and breadth of
the distribution may be reproduced by the QSO luminosity function,
taking into account the Poisson fluctuations in QSO numbers within a
\HeII-ionizing photon attenuation volume in a homogeneous IGM
\citep{2009RvMP...81.1405M}, the results found here demonstrate that a
large contribution to the breadth in the distribution is expected to
arise from variations in the \HeII\ ionizing background resulting from
the effect of inhomogeneities in the density field of the IGM on the
radiation transfer of the impinging radiation field.

The peak and FWHM in the distribution of $\eta$ at $z=2.5$ from the
simulation correspond to an effective $\Psi=\Gamma_{\rm
  HI}/\Gamma_{\rm HeII} \simeq300^{+150}_{-100}$. For a metagalactic
spectral shape $f_\nu\sim\nu^{-\alpha_{\rm MG}}$, the ratio would be
$\Gamma_{\rm HI}/\Gamma_{\rm HeII}\sim4^{1+\alpha_{\rm MG}}$. The
match to the effective $\Psi$ corresponds to effective spectral
indices of $2.9<\alpha_{\rm MG}<3.5$, much larger than the assumed
source spectral index $0.5$. The ratio corresponds to the expected
softening of the incident spectrum due to radiative transfer through
the IGM \citep{MM94, HM96}. The required rescaling of $\Psi$ imposed
on the simulations to match the measured effective \HI\ and \HeII\
optical depths corresponds to a pre-rescaled ambient radiation field
with an average $\alpha_{\rm MG}\simeq-0.25$ between the \HI\ and
\HeII\ Lyman edges arising exclusively from the radiative transfer of
the input source spectrum through the simulation volume. This
corresponds to the hardening effect of the IGM on the input radiation
field. The modification of $\Psi$ within the simulation produced by
radiative transfer is found to be highly inhomogeneous, with typically
$40-60$ per cent. variations produced in the \HeII\ ionization rate
during the passage of the \HeIII\ ionization front. The larger values
of $\Psi$ are comparable to those inferred from observations
\citep{Zheng04, Reimers05, 2006A&A...455...91F, 2010arXiv1008.2957S}.

The relation between the \HI\ and \HeII\ Doppler parameters for the
absorption systems at $z=2.5$ is shown in
Fig.~\ref{fig:bHI_bHeII_z2p5}. The \HI\ absorption lines are
restricted to those with $0.01<\tau_{0, {\rm HI}}<0.1$. In the limit
of pure thermal broadening, the \HeII\ to \HI\ ion mass ratio would
result in \HeII\ systems with $\xi=b_{\rm HeII}/ b_{\rm HI}=1/2$. In
the limit of velocity broadened lines, $\xi=1$. The upper left panels
show that the Doppler parameters lie predominantly between these two
limiting cases. Many of the points lie outside the limiting cases, a
consequence of fitting lines to blended features as well as to
saturated \HeII\ features. The wide scatter suggests the Doppler
parameter ratio $\xi$ is not a precise indicator of the relative
contributions to line broadening due to thermal motions and internal
flows. Similar results were found for selected unblended absorption
systems in the spectrum of HE~2347$-$4342 in the redshift range
$2.7<z<2.8$, although the ratios tended more towards velocity
broadening than thermal \citep{Zheng04}. No strong correlation is
found between $\xi$ and $N_{\rm HI}$ (top right panel), however the
ratio tends to increase with increasing $\eta$ (bottom right panel),
although this may be an effect of line deblending.

The ratios $\tau_{l,{\rm Ly\beta}}/ \tau_{l,{\rm Ly\alpha}}$ and
$\tau_{l,{\rm Ly\gamma}}/ \tau_{l,{\rm Ly\alpha}}$ for \HeII\ from the
{\texttt{Enzo}} simulation after reionization completes at $z\le3$ are
shown in the upper panel of Fig.~\ref{fig:tauHeIILynpLya},
renormalizing the spectra by $\tau_{l, {\rm Ly\alpha}}$ to the values
indicated. In the region of overlap, these values agree with the
simulation results quoted in \citet{2011arXiv1108.4727S}. The inferred
effective line centre optical depths $\tau_{0, {\rm Ly\alpha}}^*$
using Eq.(\ref{eq:tauLynpLyn}) are shown in the lower panel. The
resulting predicted values for $\tau_{l,{\rm Ly\gamma}}/ \tau_{l,{\rm
    Ly\alpha}}$ from Eq.(\ref{eq:tauLynpLyn}) are shown by the dotted
line in the upper panel. The values agree well with those measured
directly from the simulation, indicating the absorption is well
described by the line model, although the small deviation for large
optical depths suggests the simple one-parameter model only
approximately describes the underlying distribution at these large
optical depths.

For comparison, the value measured in the QSO HE~2347--4342 ($z_{\rm
  em}\simeq2.89$), with $\tau_{\rm Ly\alpha}^{\rm
  eff}\simeq2.393\pm0.015$ at $\langle z\rangle\simeq2.8$, is
$\tau_{\rm Ly\beta}^{\rm eff}/\tau_{\rm Ly\alpha}^{\rm
  eff}\simeq0.31\pm0.04$, using the data from
\citet{2011arXiv1108.4727S}. This agrees well with the value of
$\tau_{l,{\rm Ly\beta}}/ \tau_{l,{\rm Ly\alpha}}\simeq0.36$ predicted
from the {\texttt{Enzo}} simulation with radiative transfer, as shown
in Fig.~\ref{fig:tauHeIILynpLya}.

\section{Comparison with approximate methods}
\subsection{Approximate methods}

Two approximate simulation methods have been used in the past to
characterise the effects of \HeII\ reionization on the IGM. One method
uses $N$-body code results to represent the fluid by rescaling the
dark matter density \citep{2007MNRAS.380.1369T, 2009ApJ...694..842M};
it thus does not account for the hydrodynamical response of the gas to
heating due to radiative transfer. A similar method post-processes
hydrodynamical simulations using a radiative transfer module, but also
does not include the hydrodynamical response of the gas to the
radiative transfer \citep{BH06}. An alternative method artificially
boosts the \HeII\ heating rates in optically thin
gravity-hydrodynamics reionization simulations to mimic the boost in
temperatures resulting from radiative transfer effects
\citep{2000ApJ...534...57B}.
 
To estimate the importance of hydrodynamical feedback effects on the
observed properties of the \Lya\ forest, we compare the results of the
hydrodynamics simulations including radiative transfer with the
results of the $N$-body radiative transfer code {\texttt{PMRT}}
\citep{2007MNRAS.380.1369T}. {\texttt{PMRT}} assumes the baryons trace
the dark matter. The code is thus not able to cope with sudden
increases in the pressure forces as the gas is heated and driven out
of shallow potential wells or heated by shocks in collapsing
structures.

Past comparisons with full hydrodynamics computations without
radiative transfer show that the method recovers the distribution of
Doppler parameters to high accuracy, although the median Doppler
parameter may be slightly broadened by $1-2\kms$ \citep{MW01}. A
pseudo-hydrodynamics scheme with radiative transfer has the advantages
over a fully hydrodynamical one of ease of implementation and a much
reduced memory requirement, permitting simulations in larger boxes at
the high spatial resolution required to resolve the structures in the
IGM that give rise to the full range of measurable absorption
systems. Here we examine how successful {\texttt{PMRT}} simulations
are at capturing the essential features of the \Lya\ forest predicted
by full hydrodynamics computations including radiative transfer.

The {\texttt{PMRT}} runs used $512^3$ gravitating particles, with the
identical initial particle distributions in position and velocity as
for the {\texttt{Enzo}} runs, and the identical cosmological
parameters. It used the identical radiative transfer module as for the
{\texttt{Enzo}} runs, using $256^3$ cells to represent the
fluid. Following \citet{2007MNRAS.380.1369T}, all cells with fewer
than 100 particles were convolved over a radius of 1.5 cells to match
the gas distribution from {\texttt{Enzo}}.

We also implement the alternative method of artificially boosting the
\HeII\ heating rate in a hydrodynamical simulation in the optically
thin limit to make up for the shortfall in the additional heating due
to radiative transfer effects. A boost by a factor of 2 was found to
provide sufficient heat to broaden the absorption lines close to the
measured values at $z\simeq3$ \citep{2000ApJ...534...57B}. We
recompute the reionization simulation using {\texttt{Enzo}} with the
\HeII\ heating rate per \HeII\ ion boosted to $G_{\rm HeII}^\prime =
2G_{\rm HeII}$, and the radiative transfer switched off.

\subsection{Physical properties of IGM}

\begin{figure}
\scalebox{0.5}{\includegraphics{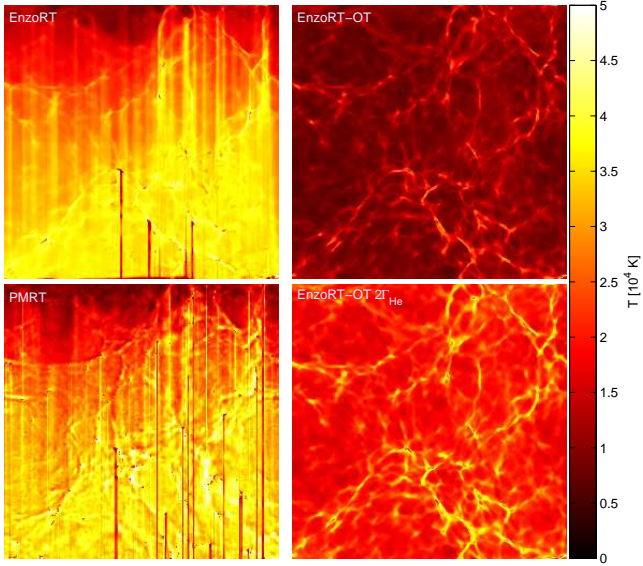}}
\caption{Temperature map at $z=3.3$, shortly before \HeII\
  reionization completes. The ionization front sweeps across the boxes
  from top to bottom. The maps are shown for {\texttt{Enzo}} with radiative
  transfer (upper left), in the optically thin limit (upper right), in
  the optically thin limit with the \HeII\ heating rate doubled (lower
  right), and for the pseudo-hydrodynamics code {\texttt{PMRT}} (lower
  left). The box side is $25h^{-1}\,$~Mpc (comoving).
}
\label{fig:Tmap-4Panel-z_3.3}
\end{figure}

\begin{figure}
\scalebox{0.5}{\includegraphics{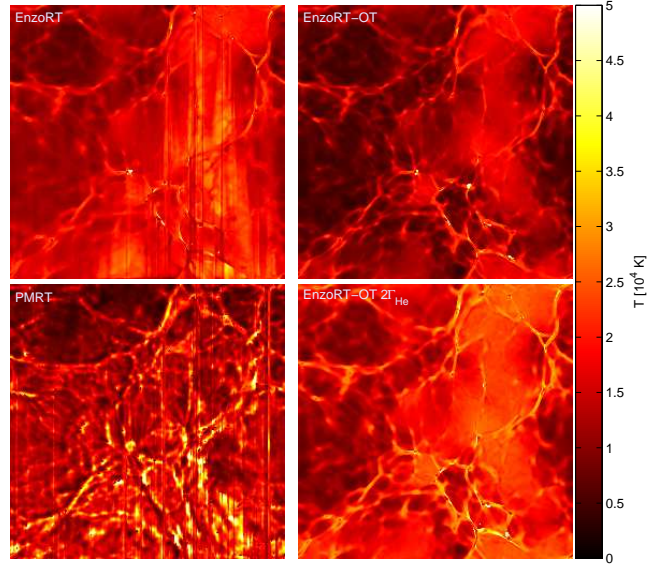}}
\caption{Temperature map at $z=2.0$, well after \HeII\
  reionization completes. The ionization front swept across the boxes
  from top to bottom. The maps are shown for {\texttt{Enzo}} with radiative
  transfer (upper left), in the optically thin limit (upper right), in
  the optically thin limit with the \HeII\ heating rate doubled (lower
  right), and for the pseudo-hydrodynamics code {\texttt{PMRT}} (lower
  left). The box side is $25h^{-1}\,$~Mpc (comoving).
}
\label{fig:Tmap-4Panel-z_2}
\end{figure}

Maps of the IGM temperature are shown in
Fig.~\ref{fig:Tmap-4Panel-z_3.3} at $z=3.3$, shortly before \HeII\
reionization completes, and in Fig.~\ref{fig:Tmap-4Panel-z_2} at
$z=2$, well after completion. At $z=3.3$, the {\texttt{PMRT}} run
matches most closely to the {\texttt{Enzo}} with RT simulation. There
are, however, some conspicuous differences. The temperature structures
tend to be sharper in the {\texttt{PMRT}} simulation, as it lacks both
shock heating and the pressure forces that smooth the gas. The effect
of shadowing by dense clumps is also more prevalent in the
{\texttt{PMRT}} simulation. This is a consequence of the inability of
the dense regions to expand due to pressure forces as the ionization
front sweeps across and boosts the gas temperature and pressure, as
discussed in Sec.~\ref{subsec:diffimp}. The {\texttt{Enzo}} simulation
with the boosted \HeII\ heating rate recovers the high temperatures in
the filaments, but the temperature in the underdense regions, although
warmer than the optically thin simulation, still lies well below the
{\texttt{Enzo}} simulation with radiative transfer.

By $z=2.0$, the {\texttt{PMRT}} simulation shows significant
discrepancies in the gas temperature within the filaments compared
with the {\texttt{Enzo}} RT simulation. In contrast to the
hydrodynamics simulation which accounts for the hydrodynamical
response of the gas to the extra heating following \HeII\
reionization, allowing it to expand and cool, the {\texttt{PMRT}}
simulation keeps the gas too hot. In underdense regions, the
{\texttt{PMRT}} simulation over-estimates the amount of adiabatic
cooling, resulting in the gas becoming too cold. The {\texttt{Enzo}}
simulation with the boosted \HeII\ heating rate is better able to
reproduce the temperatures in the underdense regions, but maintains
the gas in the filaments at too high temperatures compared with the
{\texttt{Enzo}} RT simulation.

\begin{figure}
\scalebox{0.5}{\includegraphics{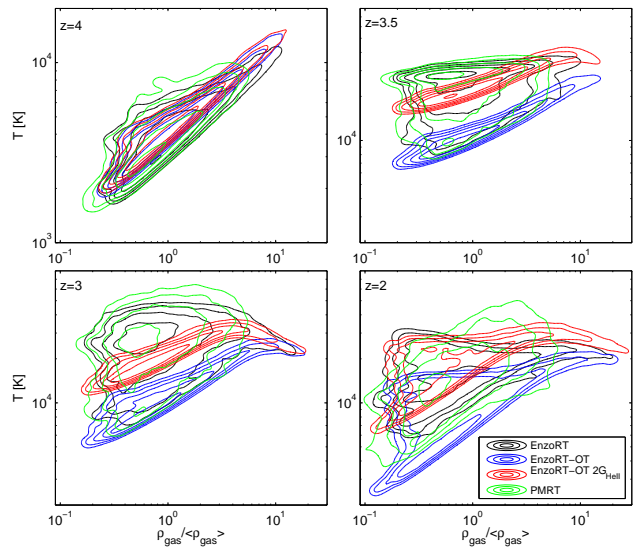}}
\caption{Temperature-density distributions at $z=4.0$, 3.5, 3.0 and
  2.0 following \HeII\ reionization, for computation using
  {\texttt{Enzo}} with radiative transfer,
  {\texttt{Enzo}} in the optically thin limit,
  {\texttt{Enzo}} in the optically thin limit but with the \HeII\
  heating rate doubled, and the
  pseudo-hydrodynamics RT code {\texttt{PMRT}}. Note
  the smaller temperature scale at $z=4.0$, prior to \HeII\
  reionization. The distributions are normalised to unity. The contour
  levels are at probability density levels (per $\diff\log_{10}
  T$-$\diff\log_{10}\rho/\langle\rho\rangle$) 0.1, 0.32, 1.0, 3.2, 10,
  32, \dots.
}
\label{fig:rho_Tdist_multiz_4models}
\end{figure}

\begin{figure}
\scalebox{0.5}{\includegraphics{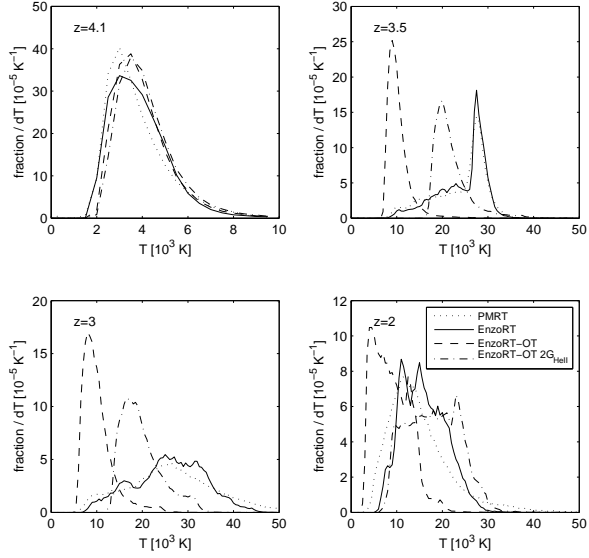}}
\caption{Temperature distributions at $z=4.1$, 3.5, 3.0 and 2.0 for
  computation using {\texttt{Enzo}} with radiative transfer (solid
  lines), {\texttt{Enzo}} in the optically thin limit (dashed lines),
  {\texttt{Enzo}} in the optically thin limit but with the
  \HeII\ heating rate doubled (dot-dashed lines), and the
  pseudo-hydrodynamics RT code {\texttt{PMRT}} (dotted lines). Note
  the smaller temperature scale at $z=4.1$, prior to
  \HeII\ reionization.
}
\label{fig:Tdist_Models}
\end{figure}

These effects may be quantified using the temperature--density
distributions extracted from the simulations. The temperature--density
distribution obtained with {\texttt{PMRT}} closely resembles that
found using {\texttt{Enzo}} with radiative transfer at $z\ge3.0$, as
shown in Fig.~\ref{fig:rho_Tdist_multiz_4models}, with reasonably good
agreement down to $z=2$. The distributions overlap very closely,
although the hydrodynamics computation shows a somewhat tighter and
flatter distribution in temperatures, especially towards $z=2$. While
the temperatures in the optically thin simulation with boosted \HeII\
heating rate agree well with the RT simulations for
$\rho/\langle\rho\rangle > 3$ over $3.0\le z\le 3.5$, the temperatures
are appreciably lower in structures with $\rho/\langle\rho\rangle <
1$. At $z=2$, good agreement is found between the {\texttt{Enzo}} RT
and boosted \HeII\ heating rate simulations for underdense structures,
but the latter produces too high temperatures for overdense
structures. The distribution from the optically thin {\texttt{Enzo}}
simulation agrees well with the others at $z=4$, but once \HeII\
reionization commences, the temperatures are systematically too low.

The temperature distributions of the {\texttt{Enzo}} with radiative
transfer and the {\texttt{PMRT}} runs are very similar, as shown in
Fig.~\ref{fig:Tdist_Models}, which may be expected from the overall
agreement in the density-temperature distrubtions. By $z=2.0$,
however, the {\texttt{PMRT}} simulation systematically underestimates
the temperatures by $3-5\times10^3$~K. Most of the volume is too cold
in the {\texttt{Enzo}} simulation with the boosted \HeII\ heating rate
for $3.0\le z \le3.5$, although a tail of warmer temperatures results
from overdense material, as shown in
Fig.~\ref{fig:rho_Tdist_multiz_4models}. By $z=2$, these temperatures
extend the distribution towards too high values compared with the RT
simulations.

\subsection{Spectral signatures}

\begin{figure}
\scalebox{0.45}{\includegraphics{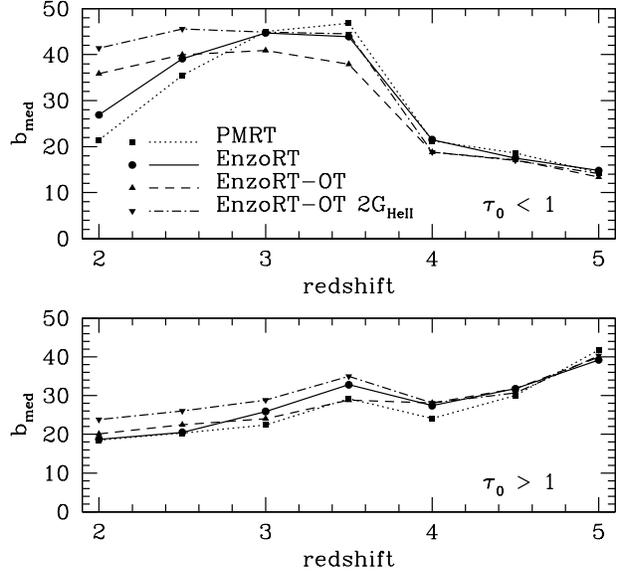}}
\caption{Evolution in the median Doppler parameter for the computation
  using {\texttt{Enzo}} with radiative transfer (solid lines),
  {\texttt{Enzo}} in the optically thin limit (dashed lines),
  {\texttt{Enzo}} in the optically thin limit with the \HeII\ heating
  rate doubled (dot-dashed lines), and the pseudo-hydrodynamics code
  {\texttt{PMRT}} (dotted lines). (Top panel)\ Results for lines
  optically thin at line centre. (Bottom panel)\ Results for lines
  optically thick at line centre.
}
\label{fig:bmed_evol}
\end{figure}

To assess the observational impact of the physical differences arising
from the hydrodynamical response of the gas compared with the
pseudo-hydrodynamics simulation on the \Lya\ forest, sample spectra
are drawn from the {\texttt{Enzo}} and {\texttt{PMRT}} simulations,
and the absorption features fit. As is clear from
Figs.~\ref{fig:pecv_map} and \ref{fig:pecv_hist}, reionization induces
substantial peculiar velocities to the gas that a pseudo-hydrodynamics
scheme like {\texttt{PMRT}} is unable to reproduce. The differences
are small compared with the peculiar velocities present due to large
scale flows, as is evident by a comparison with the full peculiar
velocity distribution shown by the rightmost curve in
Fig.~\ref{fig:pecv_hist}. The comparison between the {\texttt{Enzo}}
run with radiative transfer and the run without reionization, however,
suggests differences as great as $10\kms$ are produced, with
differences of $1-2\kms$ typical. These should be reflected in the
line widths, although the differences in gas temperature will affect
the line widths as well. For a pre-\HeII\ ionization temperature of
5000~K, a temperature boost to 20000~K will double the thermal line
width, increasing it to $18\kms$. The additional broadening due to the
increase in peculiar velocity will augment the increase due to thermal
broadening by at most only a few kilometres per second. In comparison
with the optically thin reionization simulation, the additional
peculiar velocities induced when radiative transfer is included add
negligibly to the broadening.

The evolution of the resulting median Doppler parameters is shown in
Fig.~\ref{fig:bmed_evol}. The values generally agree well for the
absorption lines optically thick at the line centre (bottom panel),
although with deviations of $\sim\pm4\kms$ from the {\texttt{Enzo}} RT
simulation. The optically thin {\texttt{Enzo}} simulations agree best
with the RT run at $z>3$, with the simulations with the boosted and
non-boosted \HeII\ heating rates bracketing the RT simulation, but
producing too broad features by $z<3$, where the {\texttt{PMRT}}
simulation gives near perfect agreement.

By contrast, the {\texttt{PMRT}} simulation recovers the median
Doppler parameters for the absorption lines optically thin at line
centre for $z>3$. These weaker lines arise from gas too rarefied to
maintain atomic thermal balance with the photoionization heating rate,
so that including the temperature heating due to radiative transfer is
required. The good agreement with the {\texttt{Enzo}} RT simulation
results shows that hydrodynamical feedback constributes little
additional broadening to the lines. The optically thin {\texttt{Enzo}}
simulation produces lines that are $\sim5\kms$ too narrow at
$3<z<3.5$. Boosting the \HeII\ heating rate recovers the median
Doppler parameter well for $3<z<3.5$, but overheats the gas at lower
redshifts. At $z<3$, adiabatic cooling in the {\texttt{PMRT}}
simulation produces features that are too narrow compared with the
{\texttt{Enzo}} RT simulation. None of the approximate methods
provides an adequate description of the absorption line widths over
the full redshift range of interest.

\section{Conclusions}

A radiative transfer module based on the method of
\citet{2007MNRAS.380.1369T} was incorporated into the
gravity-hydrodynamics code {\texttt{Enzo}} to investigate the impact
of \HeII\ reionization on the structure of the IGM. The RT module is
specifically designed to account for the heating by the hardening of
the radiation field on passing through inhomogeneities in the density
field, crucial for capturing the heating due to hard sources like
QSOs. The combined code is able to capture the hydrodynamical feedback
following the reionization of both hydrogen and helium.

A simulation of a plane-wave ionization front sweeping across a volume
$25\,h^{-1}\,{\rm Mpc}$ (comoving) on a side, representing a
characteristic $I$-front produced by photoionizing sources, was
performed. The source spectrum was initially a starburst, transforming
gradually into a QSO spectrum $f_\nu\sim\nu^{-0.5}$ over the redshift
interval $3<z<4$ to represent the onset of \HeII\ reionization by QSO
sources. Simulations in the optically thin limit, one without and one
with the \HeII\ heating rate doubled, and an $N$-body based
pseudo-hydrodynamics simulation with radiative transfer using
{\texttt{PMRT}}, are performed with identical initial conditions to
the {\texttt{Enzo}} radiative transfer simulation in order to examine
how well results from these simpler approximate methods compare.

Substantial heating is introduced by \HeII\ reionization, with
temperatures exceeding those following \HeII\ reionization in the
optically thin limit by typically $10-20\times10^3$~K. High
temperatures, in excess of $35\times10^3$~K, are produced in the
``shadows'' of overdense structures as the \HeII\ is photoionized by a
radiation field hardened on passing through the overdense
structures. While peculiar velocities are boosted by $0.1-10\kms$ by
$z=3$ over a simulation with no \HeII\ reionization, the differences
compared with a simulation with \HeII\ reionization in the optically
thin limit are typically smaller than $1\kms$. The expansion of the
gas following \HeII\ reionization, however, plays an important role in
curtailing the hardening of the radiation field by reducing the gas
density in small haloes. As a result, the temperature-density relation
is somewhat tighter and flatter compared with the corresponding
{\texttt{PMRT}} simulation, for which no hydrodynamical response is
accounted for. In rare regions, the temperature in dense regions when
radiative transfer is included is lower than the corresponding
{\texttt{Enzo}} simulation in the optically thin limit, a consequence
of complete shadowing of \HeII-ionizing photons by dense clumps in the
radiative transfer computation.

The gas flows induced by \HeII\ reionization alter the gas density
field on large scales (several comoving megaparsecs in extent), with
density reductions by as much as 10--20 per cent. compared with a
simulation without \HeII\ reionization. In much smaller regions, the
gas is pushed into structures with density increases of up to 50 per
cent. Compared with a simulation with \HeII\ reionized in the
optically thin limit, the density changes are much reduced, with large
scale reductions of $\sim5$ per cent., and density enhancements of a
few to several per cent. in complex density regions.

Because the dark matter is coupled gravitationally to the gas,
displacements in the gas will produce displacements in the dark matter
as well. Large scale coherent reductions in the dark matter density of
up to 0.5 per cent. are found at $z=3$, increasing to nearly 1 per
cent. by $z=2$. Small dense regions show enhancements in density by
0.2 per cent. at $z=3$ to 1 per cent. at $z=2$. The dark matter and
gas density displacements may produce distortions of a few per
cent. in the matter power spectrum on comoving wavenumber scales of
$k>0.5\,h{\rm Mpc}^{-1}$.

The reionization of \HeII\ has a substantial impact on the \HI\ \Lya\
forest, as would be measured in the spectra of background
QSOs. Synthetic spectra are drawn from the simulations and normalised
according to measured values of the mean \HI\ and \HeII\ transmission
values. The most conspicuous effect is on the broadening of the
absorption features. As the Doppler parameter distribution may be
measured to high accuracy in a given QSO spectrum, with a median
Doppler parameter determined to a precision of $\sim1\kms$ or better,
the line widths are a potentially powerful means of constraining the
\HeII\ reionization process and the nature of the ionizing
sources. (Alternative measures of the line shapes, such as discrete
wavelet coefficients or the flux curvature, may similarly be sensitive
probes of the reionization history.) The heating induced by radiative
transfer, however, will affect the other flux-related statistics as
well.

Compared with \HeII\ reionization in the optically thin limit, the
boost in temperatures when radiative transfer is included produces a
larger fraction of weak \HI\ absorption features with optical depths
$<0.1$, and a smaller fraction of moderate optical depth systems with
optical depths $0.1-0.6$. This may be accounted for by the increase in
the line widths and reduction in the neutral fraction at a given gas
density due to the higher temperatures, as both tend to reduce the
line centre optical depth of the absorption features. The
modifications to the optical depths result in a shift in the pixel
flux cumulative distribution, with a maximum difference between the
cumulative distributions from the {\texttt{Enzo}} runs with and
without radiative transfer of $\sim0.1$ at $z=3$, readily detectable
in measured spectra.

The \HI\ \Lya\ forest spectra are found to be fully resolvable into
absorption features. The resulting \HI\ column density distribution of
the fit lines is somewhat broader when radiative transfer is taken
into account compared with the optically thin simulation. Fewer narrow
lines are found in the simulation with radiative transfer than
without.  The median Doppler parameter evolves little between
$2.5<z<3.5$, with $b_{\rm med}\simeq35\kms$, but decreases to $b_{\rm
  med}\lsim25\kms$ by $z=2$. Dividing the absorption features into
those optically thin and optically thick at line centre reveals a
sharp difference in the evolution of the features. Those optically
thick evolve slowly, increasing from $b_{\rm med}\simeq27\kms$ when
\HeII\ reionization begins at $z=4.0$ to $b_{\rm med}\simeq33\kms$ at
$z=3.5$, then gradually diminishing to $b_{\rm med}\simeq19\kms$ at
$z=2.0$. By contrast, the median Doppler parameter of the optically
thin lines increases rapidly from $b_{\rm med}\simeq21\kms$ at $z=4.0$
to $b_{\rm med}\simeq44-45\kms$ at $3.0\le z \le 3.5$, then diminishes
slightly to $b_{\rm med}\simeq39\kms$ at $z=2.5$ and rapidly to
$b_{\rm med}\simeq27\kms$ by $z=2.0$. The median Doppler widths of the
optically thick systems agree well with those measured. By contrast,
the median values of the optically thin systems at $3<z<3.5$
substantially exceed those measured, suggesting either \HeII\
reionization was imposed too late in the simulation, having just
completed at $z\simeq3$, or that the source spectrum is too hard. A
previous {\texttt{PMRT}} simulation with \HeII\ reionization initiated
at $z=5.0$ \citep{2007MNRAS.380.1369T} produced a substantially lower
peak temperature of $15\times10^3$~K by $z=3.0$, cooler than found
here by $10-20\times10^3$~K. Cooler temperatures may thus be achieved
if \HeII\ reionization were substantially completed prior to $z=3$,
although this may be at tension with recent claims that \HeII\
reionization extended to $z<3$. The {\texttt{PMRT}} simulations also
found cooler temperatures arose when a single power-law spectrum
reionized both hydrogen and helium, so that the temperatures are
sensitive to the amount of \HeII\ reionization during the hydrogen
reionization epoch. Clearly simulations allowing for a more complex
\HeII\ reionization history involving a range of source histories and
spectra are required to fully match the data.

The {\texttt{PMRT}} simulation matches the median Doppler width of the
optically thick absorbers in the {\texttt{Enzo}} radiative transfer
simulation to an accuracy of $2\kms$ at $z\ge 4$ and better than
$0.5\kms$ at $2\le z \le 2.5$, but underpredicts the widths by as much
as $3.5\kms$ at intermediate redshifts. For the absorbers optically
thin at line centre, the agreement is to better than $2\kms$ for
$3.0\le z \le 5$, but underpredicts the widths by $\sim5\kms$ by
$z=2.0$.

Doubling the \HeII\ heating rate in an optically thin {\texttt{Enzo}}
simulation produces a close match to the median Doppler parameter of
the optically thin absorbers from the {\texttt{Enzo}} simulation with
radiative transfer over $3.0\le z \le 3.5$, agreeing to within
$0.6\kms$. But by $z=2.0$, the median Doppler parameter is
overpredicted by $\sim15\kms$. The Doppler parameters of the
absorption systems optically thick at line centre are overpredicted by
2-$6\kms$ over $2.0\le z \le 3.5$. The approximate methods are thus
unable to reproduce the correct amount of broadening to the accuracy
with which it is measured over the full range of redshifts for which
the forest may be observed in high resolution, high signal-to-noise
ratio spectra.

Just as absorption features may fully account for the pixel flux
distribution in \Lya, a line blanketing model accounts well for the
evolution in the mean transparency of the IGM in higher order Lyman
series lines. Effective optical depth ratios for the \HI\ of
$\tau_{\rm Ly\beta}^{\rm eff}/ \tau_{\rm Ly\alpha}^{\rm
  eff}\simeq0.36$ and $\tau_{\rm Ly\gamma}^{\rm eff}/ \tau_{\rm
  Ly\alpha}^{\rm eff}\simeq0.21$ are predicted at $z=3.0$ from the
{\texttt{Enzo}} radiative transfer simulation.

Radiative transfer through density inhomogeneities in the IGM results
in a wide spread of \HeII\ \Lya\ optical depths compared with \HI\
even after \HeII\ reionization has completed in the simulation at
$z=3$. A broad peak centred at $\tau_{\rm HeII}/\tau_{\rm
  HI}\simeq105$ is found at $z=3.0$, diminishing to 80 at $z=2.5$ and
75 at $z=2.0$.

As for the \HI, the pixel flux distribution for \HeII\ \Lya\ may be
fully accounted for by absorption lines for $2.0\le z \le 2.5$. At
$z=3.0$, however, the mean transmission is too low for meaningful
absorption line fitting to be performed. Matching to the corresponding
\HI\ absorption features results in a broad distribution of the column
density ratio $\eta=N_{\rm HeII}/ N_{\rm HI}$, with a peak of
$\eta\simeq130$ at $z=2.5$, well matching, though somewhat narrower
than, the distribution measured in the ultraviolet spectra of high
redshifts QSOs. While the value for $\eta$ at the distribution peak
results from the rescaling of the relative \HI\ and \HeII\ ionization
rates required to match the measured effective \HI\ and \HeII\ \Lya\
optical depths, the distribution width is produced by the simulation
and suggests that much of the observed spread may arise naturally from
the radiative transfer through an inhomogeneous IGM.

Comparison between the column density and optical depth ratios
indicates a typical Doppler parameter ratio of $\xi=b_{\rm HeII}/
b_{\rm HI}\simeq0.4$, close to the thermally broadened limit of
$\xi=0.5$. A comparison based on well deblended, unsaturated systems
with line centre \HI\ optical depths $0.01<\tau_0<0.1$ shows a range
in Doppler parameter ratios peaking between $0.5<\xi<1.0$, the latter
corresponding to the limit of velocity-broadened features.

Line blanketing accounts well for the higher order \HeII\ Lyman
transmission through the IGM as well. The simulations predict
effective optical depth ratios of $\tau_{\rm Ly\beta}^{\rm eff}/
\tau_{\rm Ly\alpha}^{\rm eff}\simeq0.36$ at $z=2.5$, in agreement with
the measured value, and $\tau_{\rm Ly\gamma}^{\rm eff}/ \tau_{\rm
  Ly\alpha}^{\rm eff}\simeq0.20$ (currently unmeasured).

A range of models, allowing for multiple QSO sources turning on over a
range of redshifts with a variety of luminosities, spectral shapes and
lifetimes, would be required to accurately match the \HI\ and \HeII\
absorption data as measured in the spectra of high redshift background
QSOs. The simulations presented here demonstrate that \HeII\
reionization simulations including radiative transfer are required to
fully exploit the growing body of high quality IGM data at both
optical and ultraviolet wavelengths.

\section*{Acknowledgments}
The computations reported here were performed using the SUPA
Astrophysical HPC facility and facilities funded by an STFC
Rolling-Grant. E.T. is supported by an STFC Rolling-Grant.
Computations described in this work were performed using the
{\texttt{Enzo}} code developed by the Laboratory for Computational
Astrophysics at the University of California in San Diego
(http://lca.ucsd.edu).

\bibliographystyle{mn2e-eprint}
\bibliography{apj-jour,ms}
\label{lastpage}

\end{document}